\documentstyle[onecolumn]{mn}

\input epsf.sty

\def\upi{\pi}

\def\etal{{et al.~}}
\def\fr#1#2{{\textstyle {#1 \over #2}}}

\title[Scale-free dynamical models for galaxies]{Scale-free dynamical 
       models for galaxies:\\ flattened densities in spherical potentials}

\author[Jos H.~J.~de Bruijne, Roeland P.~van der Marel and P.~Tim de Zeeuw]
       {Jos~H.~J.~de~Bruijne,$^1$ Roeland~P.~van~der~Marel$^{2,3}$ 
            and P.~Tim~de~Zeeuw$^1$\\
        $^1$Sterrewacht Leiden, Postbus 9513, 2300 RA Leiden, the
              Netherlands\\ 
        $^2$Institute for Advanced Study, Princeton, NJ 08540, USA\\ 
        $^3$Hubble Fellow}

\journal{submitted to Mon.~Not.~Royal Astr.~Soc.~on Dec 19, 1995.}

\begin{document}

\maketitle

\begin{abstract}
This paper presents two families of phase-space distribution functions
that generate scale-free {\it spheroidal\/} mass densities in
scale-free {\it spherical\/} potentials. The assumption of a spherical
potential has the advantage that all integrals of motion are known
explicitly. The `case~I' distribution functions are anisotropic
generalizations of the flattened $f(E,L_z)$ model, which they include
as a special case. The `case~II' distribution functions generate
flattened constant-anisotropy models. Free parameters control the
radial power-law slopes of the mass density and potential, the
flattening of the mass distribution, and the velocity dispersion
anisotropy. The models can describe the outer parts of galaxies and
the density cusp structure near a central black hole, but also provide
general insight into the dynamical properties of flattened
systems. Because of their simplicity they provide a useful
complementary approach to the construction of flattened
self-consistent three-integral models for elliptical galaxies.

The dependence of the intrinsic and projected properties on the model
parameters and the inclination is described. The case~I models have a
larger ratio of rms tangential to radial motion in the equatorial
plane than on the symmetry axis, the more so for smaller axial ratios.
The case~II models have a constant ratio of rms tangential to radial
motion throughout the system, as characterized by Binney's parameter
$\beta$. The maximum possible ratio $v_{\rm p} / {\sigma}_{\rm p}$ of
the mean projected line-of-sight velocity and velocity dispersion on
the projected major axis always decreases with increasing radial
anisotropy. The observed ratio of the rms projected line-of-sight
velocities on the projected major and minor axes of elliptical
galaxies is best fit by the case~II models with $\beta \ga 0$. These
models also predict non-Gaussian velocity profile shapes consistent
with existing observations. 

The distribution functions are used to model the galaxies NGC 2434
(E1) and NGC 3706 (E4), for which stellar kinematical measurements out
to two effective radii indicate the presence of dark halos (Carollo et
al.). The velocity profile shapes of both galaxies can be well fit by
radially anisotropic case~II models with a {\it spherical\/}
logarithmic potential. This contrasts with the $f(E,L_z)$ models
studied previously, which require {\it flattened} dark halos to fit
the data.
\end{abstract}

\begin{keywords}
galaxies: elliptical and lenticular, cD
-- galaxies: individual: NGC 2434 and NGC 3706
-- galaxies: kinematics and dynamics
-- galaxies: structure
-- line: profiles.
\end{keywords}


\section{Introduction}

Elliptical galaxies are dynamically complex systems (e.g., de Zeeuw \&
Franx 1991). Many unsolved problems regarding their structure still
exist. Studies of the presence and properties of dark halos and
massive central black holes have been hampered by lack of information
about the stellar velocity dispersion anisotropy.  However, the body
of observational data from which such knowledge can be derived (at
least in principle) is growing steadily. In particular, deviations of
the shapes of the stellar line-of-sight velocity distributions, or
`velocity profiles' (VPs), from Gaussians can now be measured reliably
(e.g., van der Marel \etal 1994a; Bender, Saglia \& Gerhard 1994;
Carollo \etal 1995). Detailed dynamical models based on phase-space
distribution functions (DFs) are needed to interpret such data.

Very few, if any, galaxies are spherical. For a proper interpretation
of the high-quality data that can now be obtained, flattened
axisymmetric models appear a minimum requirement. Axisymmetric
potentials admit two classical integrals of motion, the energy per
unit mass $E$, and the angular momentum component per unit mass
parallel to the symmetry axis, $L_z$. In addition, there is usually a
non-classical third integral, $I_3$ (Binney \& Tremaine 1987). The DF
$f$ is generally a function $f(E,L_z,I_3)$. A subset of all DFs is
formed by those that depend only on the two classical integrals, i.e.,
$f(E,L_z)$. Construction of such models has only recently become
practical, and as a result they are now being studied in great detail
(e.g., Hunter \& Qian 1993; Evans 1993, 1994; Dehnen \& Gerhard 1994;
Evans \& de Zeeuw 1994; Qian \etal 1995; Dehnen 1995; Kuijken 1995;
Magorrian 1995). These models properly take flattening into account,
and are very useful for their relative simplicity. However, they are
special in that the radial and vertical velocity dispersions are
everywhere equal (i.e., $\sigma_R \equiv \sigma_z$). This is generally
not the case in real galaxies (Binney, Davies \& Illingworth 1990; van
der Marel 1991). For most practical applications (e.g., to demonstrate
that no model without a black hole or dark halo can fit a particular
set of observations) one needs to construct more general models with
three-integral DFs. This is complicated, because $I_3$ is generally
not known analytically, and may not even exist for all orbits.
Dejonghe \& de Zeeuw (1988) considered axisymmetric models with a
potential of St\"ackel form, in which a global third integral {\it
is\/} known analytically. Dehnen \& Gerhard (1993) considered
axisymmetric models with a flattened isochrone potential in which the
third integral can be approximated using Hamiltonian perturbation
theory. Models of this kind are currently being applied to fit the
kinematics of real galaxies (e.g., Dejonghe \etal 1995; Matthias \&
Gerhard 1995).  An alternative to these semi-analytical methods is a
completely numerical approach in which individual orbits are
superimposed (as in Schwarzschild 1979) using a linear programming,
maximum entropy, non-negative least squares, or related technique.
Several groups are now working on this. All these methods require
substantial analytic or numerical effort.

A case of `intermediate complexity' which has not been much studied in
the literature is that of flattened mass distributions in spherical
potentials. The assumption of a spherical potential has the advantage
that the squared angular momentum per unit mass, $L^2$, is an
explicitly known third integral, so that $f=f(E, L^2, L_z)$. White
(1985) already determined simple DF components in the spherical
logarithmic potential (see also de Zeeuw, Evans \& Schwarzschild
1995), while Kochanek (1994) solved the Jeans equations for realistic
flattened density distributions in this potential. Mathieu, Dejonghe
\& Hui (1995) constructed triaxial mass models in a spherical potential 
for the galaxy Cen A. In the present paper a detailed study is
presented of DFs for scale-free axisymmetric mass densities embedded
in scale-free spherical potentials. The general form of the DFs of
such models is derived, and two particular families of DFs are
discussed in detail. For these families most physically and
observationally interesting quantities can be determined analytically.
The models therefore allow a detailed study of the dependence of the
observable kinematical quantities on the various model parameters,
such as the power-law slopes of the mass density and potential, the
axial ratio of the density distribution, the inclination angle of the
symmetry axis with respect to the line of sight, the intrinsic
velocity dispersion anisotropy, and the amount of mean streaming. The
models provide significant insight into the dynamical structure of
flattened galaxies, and provide a useful complementary approach to the
construction of fully self-consistent models.

As an application, the issue of the evidence for massive dark halos
around elliptical galaxies is considered. Both tangential anisotropy
and the presence of a dark halo can cause the observed velocity
dispersion to remain roughly constant out to well beyond the effective
radius.  Hence, a flat observed dispersion profile does not prove the
existence of a dark halo. Carollo \etal (1995) presented observations
of the major axis kinematics and VP shapes for four elliptical
galaxies, and constructed $f(E,L_z)$ models to interpret their
data. Here we restrict the discussion to two of the four galaxies, NGC
2434 and 3706, for which Carollo \etal concluded from the observed VP
shapes that dark halos must be present. They showed that the dark
halos must be flattened, if the observed VPs are to be fit with an
$f=f(E,L_z)$ DF. We use our models to determine whether the dark halos
of these galaxies must indeed be flattened, or whether the VPs can be
fit as well with a spherical dark halo and a DF of the form
$f(E,L^2,L_z)$.

Section~2 discusses the DFs of the models, and the calculation of the
intrinsic and projected velocity moments and VP shape parameters. This
section is technical, and readers mostly interested in applications of
the models might wish to skip to Section~3 after
Section~2.1. Section~3 gives a general discussion of the various
properties of the models. The application to the Carollo \etal data is
described in Section~4. Section~5 summarizes the main results.
Appendix~A presents an algorithm for the reconstruction of a VP from
its moments.

\section{The models}

\subsection{Potential and mass density}

Throughout this paper $(r,\theta,\phi)$ denote the usual spherical
coordinates, and $(R,\phi,z)$ the usual cylindrical coordinates, with
$z$ along the symmetry axis of the mass density. The {\it relative\/}
potential $\Psi$ and the {\it relative\/} energy per unit mass ${\cal
E}=\Psi-{{1}\over{2}}v^2$ are defined as in Binney \& Tremaine
(1987). The potential decreases outwards. Its value at infinity can
either be finite, in which case one can set ${\Psi}_{\infty}=0$
without loss of generality, or it can be ${\Psi}_{\infty} =
-{\infty}$. The quantity ${\cal E}$ is the binding energy per unit
mass of a star. Only stars with ${\cal E} > {\Psi}_{\infty}$ are bound
to the system.

Scale-free spherical potentials $\Psi$ are considered, of the form
\begin{equation}
  \Psi(r) \equiv V_0^2 \> \times \> \left\{ \begin{array}{ll}
       - \ln (r/b)                    , & \qquad \delta=0          ; \\
       {1\over\delta}(r/b)^{-\delta}  , & \qquad 0 < \delta \leq 1 ,
                 \end{array} \right.  
\end{equation}
where $b$ is a reference length. The free parameter $\delta$, with $0
\leq \delta \leq 1$, determines the radial slope of the potential. The scale 
velocity $V_0$ is equal to the circular velocity at the reference
length. For $\delta=0$ the potential is the logarithmic potential, and
the circular velocity is independent of radius. For $\delta=1$ the
potential is Keplerian. If the Kepler potential is generated by a
total mass $M$, then $V_0^2 = GM/b$, with $G$ the universal constant
of gravitation.

Mass densities are considered that are power laws on oblate spheroids:
\begin{equation}
  \rho (R,z) \equiv {\rho}_{0} \> 
        ({R^2 \over b^2} + {{z^2} \over {b^2 \> q^2}})^{-\gamma / 2}  ,
\label{def.rho_units}
\end{equation}
where $\rho_0$ is a reference mass density, $\gamma \geq 0$ is a
constant that determines the radial fall off, and $q \leq 1$ is the
constant axial ratio of the similar concentric isodensity surfaces of
the mass distribution. The eccentricity is $e \equiv
\sqrt{1- q^2}$. The limit $q=1$, or $e=0$, describes the spherical
power law $\rho (r)= \rho_0 \, (r/b)^{-\gamma}$. Mass distributions of
the form (\ref{def.rho_units}) always produce systems with infinite
total mass: for $0 \leq \gamma < 3$, the total mass diverges at large
radii, for $\gamma = 3$, the total mass diverges at both small and
large radii, while for $\gamma > 3$, the total mass diverges at small
radii. Nonetheless, the models meaningfully describe the properties of
realistic finite-mass systems, but only at those radii where the mass
density can be approximated by equation~(\ref{def.rho_units}).

Dimensionless quantities are used throughout the remainder of this
paper: ${\tilde r} \equiv r/b$; ${\tilde R} \equiv R / b$; ${\tilde z}
\equiv z / b$; ${\tilde v} \equiv v / \sqrt{2} V_0$; ${\tilde L}
\equiv L / \sqrt{2} b V_0$; ${\tilde {\cal E}} \equiv {\cal E} /
V_0^2$; ${\tilde \Psi} \equiv \Psi / V_0^2$; ${\tilde \rho}
\equiv \rho / \rho_0$; and ${\tilde f} \equiv f / \rho_0
(2V_0^2)^{-3/2}$. Henceforth, all tildes are omitted. The potential
and mass density of the models are thus:
\begin{equation}
  \Psi(r) \equiv \left\{ \begin{array}{ll}
       - \ln r                    , & \qquad \delta=0          ; \\
       {1\over\delta} r^{-\delta} , & \qquad 0 < \delta \leq 1 ,
                 \end{array} \right.  
\label{def.Psi}
\end{equation}
and
\begin{equation}
  \rho (R,z) \> = \> (R^2 + {{z^2} \over {q^2}})^{-\gamma / 2} 
             \> = \> q^{\gamma} r^{-\gamma} 
                     (1 - e^2 \sin^2 \theta)^{-\gamma /2} .
\label{def.rho}
\end{equation}
The latter expression can be expanded in a power series in
$e^2{\sin}^2\theta$ using the binomial theorem, with the result:
\begin{equation}
  \rho (r,\theta) \> = \> q^{\gamma} r^{-\gamma} \sum_{k=0}^{\infty}
                    {{1}\over{k!}} \> (\gamma/2)_{k} \> 
                    (e^2 \sin^2\theta)^k ,
\label{rho_desired} 
\end{equation}
where $(\ldots)_{k}$ is Pochhammer's symbol, which is defined in terms
of Gamma--functions as $(x)_t \equiv \Gamma(x+t) / \Gamma(x)$ (cf.,
e.g., Gradshteyn \& Ryzhik 1994).

\subsection{Distribution functions}

\subsubsection{Self--similarity}

DFs of the form $f({\cal E}, L^2, L_z)$ are sought, that generate the
mass density~(\ref{def.rho}) in the potential~(\ref{def.Psi}). The
integrals of motion are ${\cal E} = \Psi-v^2$ (the usual factor $1/2$
in the kinetic energy term has been absorbed in the units), $L^2 = r^2
(v_{\theta}^2 + v_{\phi}^2)$ and $L_z = R v_{\phi}$. We consider first
the part of the DF that is even in $L_z$, $f_{\rm e} ({\cal E}, L^2,
L_z^2)$. This part determines the mass density completely, because the
latter is independent of a star's sense of rotation around the
symmetry axis.

The maximum angular momentum $L_{\rm max} ({\cal E})$ at a given
energy is attained by stars on circular orbits in the equatorial
plane. The squared circular velocity is $V_c^2(r) = r^{-\delta}/2$,
and hence
\begin{equation}
L_{\rm max}^2 ({\cal E}) = \left\{ \begin{array}{ll}
     {1\over2} \> \exp (-2{\cal E} - 1)
        , & \qquad \delta=0 ; \\
     {1\over2} \> [2 \delta {\cal E} / (2-\delta)]^{(\delta-2)/\delta} 
        , & \qquad 0 < \delta \leq 1 .
                           \end{array} \right.
\label{def.L_max}
\end{equation}
Without loss of generality, the DF can be considered to be a function
$f_{\rm e}({\cal E},{\zeta}^2,{\eta}^2)$, with 
\begin{equation}
  {\zeta}^2 \equiv L^2 / L_{\rm max}^2({\cal E}) , \qquad
  {\eta}^2 \equiv L_z^2 / L_{\rm max}^2({\cal E}),
\label{def.zeta_eta}
\end{equation}
so that $0 \leq \eta^2 \leq \zeta^2 \leq 1$. 

The discussion can be restricted to DFs that are {\it
self--similar\/}, since both the potential and the mass density are
scale free. For such DFs there exist constants $c_1$ and $c_2$ such
that
\begin{equation}
f_{\rm e} \, (p{\bmath r}, p^{c_1}{\bmath v}) =
   p^{c_2} f_{\rm e} \, ({\bmath r},{\bmath v}),
\end{equation}
for all ${\bmath r}$, ${\bmath v}$ and $p$. Following White (1985),
we substitute $f_{\rm e}({\cal E},{\zeta}^2,{\eta}^2)$, differentiate
with respect to $p$, and then set $p=1$. This shows that $c_1 =
-\delta / 2$ for all $0 \leq \delta \leq 1$, and that $f_{\rm e}$ must
have the general form
\begin{equation}
  f_{\rm e}({\cal E},{\zeta}^2,{\eta}^2) = F(\zeta^2, \eta^2) 
     \times \left\{ \begin{array}{ll}
            \exp (c_2 {\cal E})            ,  & \qquad \delta = 0 ; \\
            (\delta {\cal E})^{c_2/\delta} , & \qquad 0 < \delta \leq 1 ,
               \end{array}\right.
\label{f_selfsim} 
\end{equation}
where $F$ is an arbitrary non-negative function. 

\subsubsection{DF components}

The mass density is the integral of the DF over velocity space,
\begin{equation}
\rho = \int {\rm d}^3{\bmath v} \> f =
  \int_{0}^{\sqrt{\Psi - \Psi_{\infty}}} {\rm d}v \> v^2 
  \int_{0}^{\upi} {\rm d}\xi \> \sin\xi 
  \int_{0}^{2\upi} {\rm d}\tau \> f_{\rm e}
\label{rhointf} ,
\end{equation}
where the variables $(v,\xi,\tau)$ are spherical coordinates in
velocity space: 
\begin{equation}
  v_{r} = v\cos\xi, \qquad
  v_{\theta} = v\sin\xi\cos\tau, \qquad
  v_{\phi} = v\sin\xi\sin\tau.
\label{corvelsph}
\end{equation}
It is convenient to transform to the new integration variables
\begin{equation}
  \overline{{\cal E}} \equiv {\cal E}/\Psi = 1 - (v^2/\Psi) , \qquad
  \overline{{\zeta}^2} \equiv \sin^2 \xi , \qquad
  \overline{{\eta}^2} \equiv \sin^2\tau.  
\label{corvelnew}
\end{equation}
This results in 
\begin{eqnarray}
  \rho = \Psi^{3/2} 
         \int_{\Psi_{\infty} / \Psi}^{1} {\rm d}\overline{{\cal E}} \>
            (1-\overline{{\cal E}})^{1/2} \> 
         \int_{0}^{1} {\rm d}\overline{{\zeta}^2} \>
            (1-\overline{{\zeta}^2})^{-1/2} \>
         \int_{0}^{1} {\rm d}\overline{{\eta}^2} \>
            (1-\overline{{\eta}^2})^{-1/2} \>
            (\overline{{\eta}^2})^{-1/2} \> f_{\rm e} .
\label{rhointf2}
\end{eqnarray}
The lower limit of the outermost integral is $\Psi_{\infty} / \Psi =
0$ for $0 < \delta \leq 1$. For the logarithmic potential $\delta=0$
it is $\Psi_{\infty} / \Psi = -\infty$.

The integrals of motion $({\cal E},{\zeta}^2,{\eta}^2)$ can be expressed in
terms of the integration variables $(\overline{{\cal E}},
\overline{{\zeta}^2}, \overline{{\eta}^2})$ as follows:
\begin{equation}
   {\cal E} = \Psi \overline{{\cal E}} , 
\qquad
   {\eta}^2 = \sin^2\theta \> \overline{{\eta}^2} \zeta^2 , 
\qquad
   {\zeta}^2 =
      \Psi r^2 \overline{{\zeta}^2} \> (1-\overline{{\cal E}}) \> \times \>
      \left\{ \begin{array}{ll}
         2 \> \exp(2 \Psi \overline{{\cal E}} + 1)
              , & \qquad \delta=0 ; \\
         2 \> [2 \delta \Psi \overline{{\cal E}} / 
              (2-\delta)]^{(2-\delta)/\delta}
              , & \qquad 0 < \delta \leq 1 .
      \end{array} \right.
\label{Int_velnew}
\end{equation}
We restrict ourselves in equation~(\ref{f_selfsim}) to smooth
functions $F$ that can be expanded as sums of terms of the form
$\zeta^{-2\mu} \> \eta^{2\lambda}$. The entire DF is then a sum of
self-similar components of the form
\begin{equation}
  f^{[c_2,\mu,\lambda]} ({\cal E},\zeta^2,\eta^2) \equiv
      \zeta^{-2\mu} \> \eta^{2\lambda} \>
      \times \left\{ \begin{array}{ll}
           \exp(c_2 {\cal E})             ,    & \qquad \delta = 0 ; \\
           (\delta {\cal E})^{c_2/\delta} , & \qquad 0 < \delta \leq 1 .
               \end{array}\right.
\label{def.DF_comp} 
\end{equation}
Substitution of equations~(\ref{Int_velnew}) and~(\ref{def.DF_comp}) in
equation~(\ref{rhointf2}), and carrying out the triple integration,
shows that each component generates a mass density
\begin{equation}
  {\rho}^{[c_2,\mu,\lambda]} (r,\theta) \equiv 
    \int {\rm d}^3{\bmath v} \> f^{[c_2,\mu,\lambda]} = 
       D^{[c_2,\mu,\lambda]} \> r^{-c_2-(3\delta/2)} 
       \> \sin^{2\lambda}\theta
  \label{rho_component} ,
\end{equation}
where the factors $D^{[c_2,\mu,\lambda]}$ are given by\newline
\vbox{
\begin{eqnarray}
\lefteqn{
  D^{[c_2,\mu,\lambda]} =
    {\cal B} (\fr12,  \lambda+\fr12) \>
    {\cal B} (\fr12, 1-\mu+\lambda) }\nonumber \\ 
  & & \qquad\qquad\qquad \times \> \left\{ \begin{array}{ll}
    2^{\lambda-\mu} \> \exp(\lambda-\mu) \>
    \Gamma ({3\over2}\!-\!\mu\!+\!\lambda) \>
    (c_2\!+\!{{3\delta}\over2}\!-\!2\mu\!+\!2\lambda)^{-{3\over2}+\mu-\lambda},
 & \qquad \delta = 0 ; \\
    \delta^{-3/2} \>
    \left[({\delta\over2}) ({{2-\delta}\over 2})^{{2-\delta}\over{\delta}}
         \right]^{\mu-\lambda} \>
    {\cal B} ({3\over2}\!-\!\mu\!+\!\lambda,{1\over\delta} 
             [c_2\!+\!{{3\delta}\over2}\!-\!(2\!-\!\delta) 
             (\mu\!-\!\lambda)]\!-\!{1\over2} ) ,
  & \qquad 0 < \delta \leq 1 . \\
       \end{array} \right.
\label{def.Dfac}
\end{eqnarray}
}
The function ${\cal B}(\ldots,\ldots)$ is the Beta--function, which is
defined in terms of Gamma--functions as ${\cal B}(x,y) \equiv
\Gamma(x) \Gamma(y) / \Gamma(x+y)$ (cf., e.g., Gradshteyn \& Ryzhik 1994).
The $D^{[c_2,\mu,\lambda]}$ are continuous functions of $\delta$ in
the limit $\delta \downarrow 0$. These results were obtained
independently by Evans (1995, priv.~comm.).

Equation~(\ref{rho_component}) shows that in order to reproduce the
mass density (\ref{rho_desired}) with DFs of the form
\begin{equation}
f_{\rm e}({\cal E}, \zeta^2, \eta^2) =
          \sum_\mu \sum_\lambda  \alpha^{[c_2,\mu,\lambda]} 
           f_{\rm e}^{[c_2,\mu,\lambda]} ({\cal E},\zeta^2,\eta^2), 
\label{def.superposition_DF} 
\end{equation}
one requires that 
\begin{equation}
  c_2 = \gamma - (3\delta / 2),
\qquad\qquad
  {\rm and}
\qquad\qquad
  \sum_\mu \alpha^{[c_2,\mu,k]} D^{[c_2,\mu,k]} = 
     q^\gamma \> {{1}\over{k!}} \> (\gamma/2)_{k} \> e^{2k} ,
\qquad
  ({\rm for} \>\> k=0,1,2,\ldots) .
\label{def.superposition_rho} 
\end{equation}
In addition, the expansion coefficients $\alpha^{[c_2,\mu,\lambda]}$
must be zero for all $\lambda \not= k$ (with $k = 0,1,2,\ldots$). The
value of $c_2$ ensures that the density components fall as
$r^{-\gamma}$, for all $0 \leq \delta \leq 1$. Henceforth, $c_2 =
\gamma - (3\delta/2)$ is substituted in all equations that involve~$c_2$.

\begin{table*}
\renewcommand{\arraystretch}{1.55}
\begin{minipage}{450truept}
\caption{Special cases of the function $h(x^2)$ defined in 
         equation~(26), and of the function $j(x^2)$ for $\delta=1$, 
         defined in equation~(32). The function $[j(x^2)]_{\delta=1}$
         for $\gamma=4$ can be reduced to an elementary expression 
         only when $2\beta$ is an integer.}
\begin{tabular*}{450truept}{l@{\extracolsep{\fill}} l@{\extracolsep{\fill}} l}
\hline
\noalign{\smallskip}
$\gamma$  & $h(x^2) = {}_2F_1 (1, \fr{\gamma}{2}; \fr12; x^2)$     
          & $[j(x^2)]_{\delta=1} = {}_4F_3 (1, \fr{\gamma}{2}, 
                   \fr{\gamma -2\beta +1}{2},
                   \fr{\gamma -2\beta+2}{2}; \fr12,
                   1-\beta, \gamma-\beta-\fr12; x^2)$ \\ \\
1         & ${1 \over 1-x^2}$
          & ${1-2\beta + (1+2\beta)x^2 \over (1-2\beta) (1-x^2)^2}$ \\
2         & ${1 \over 1-x^2} + {x \arcsin x \over (1-x^2)^{3/2}}$
          & ${2-2\beta +(1+2\beta)x^2 \over 2(1-\beta)(1-x^2)^2} 
             +{(3-2\beta+2\beta x^2) x \arcsin x \over 
              2(1-\beta) (1-x^2)^{5/2}}$ \\
3         & ${1+x^2 \over (1-x^2)^2}$
          & ${1-\beta +3x^2 +\beta x^4 \over (1-\beta)(1-x^2)^3}$ \\
4         & ${2+x^2 \over 2(1-x^2)^2} + {3x \arcsin x \over 2(1-x^2)^{5/2}}$
          & ${{5-2\beta} \over {(2-2\beta)(4-2\beta)}} x^{2\beta-1} 
             \fr{{\rm d}}{{\rm d}x} \left \lbrace \fr{1}{x}
             \fr{{\rm d}}{{\rm d}x} \left [ \fr{1}{x}
             \int x^{4-2\beta} h(x^2) \> {\rm d} x 
               \right ] \right \rbrace$ \\
\noalign{\smallskip}
\hline
\end{tabular*}
\end{minipage}
\renewcommand{\arraystretch}{1.0}
\end{table*}

\subsubsection{Two families of DFs}

Many DFs can be constructed that satisfy
equation~(\ref{def.superposition_rho}). Two particular cases are
discussed here, which differ in the choice of components
$f_e^{[c_2,\mu,k]}$. The first set has $\mu$ equal to the same
constant for all components, so that $f_e$ is a separable function of
$E$, $L^2/L^2_{\rm max}(E)$ and $L_z^2/L^2_{\rm max}(E)$. In the
second set, the components are chosen such that $f_e$ is a separable
function of $E$, $L^2/L^2_{\rm max}(E)$ and $L_z^2/L^2$. The latter
models turn out to have a velocity distribution anisotropy that is
independent of position (cf.~Section~2.3 below).

In case~I the DF is built entirely with components $f_{\rm
e}^{[c_2,\mu,k]}$ for which $\mu$ is equal to a constant $\beta$. The
DF is then:
\begin{equation}
  f_{\rm e}^{\rm I} ({\cal E},\zeta^2,\eta^2) = 
        C_0 \> g({\cal E}) \> \zeta^{-2\beta} \> j(e^2 \eta^2) ,
\label{dfI_even}
\end{equation}
where the functions $j$ and $g$ are defined as
\begin{equation}
  j(e^2 \eta^2) \equiv \sum_{k=0}^{\infty} a_k \> (e^2 \eta^2)^k ,
\qquad\qquad
  g({\cal E}) \> \equiv \> \left\{ \begin{array}{ll}
           \exp(\gamma {\cal E}) , & \qquad \delta = 0 ; \\
           (\delta {\cal E})^{(\gamma/\delta)-(3/2)} 
                                 , & \qquad 0 < \delta \leq 1 ,
               \end{array}\right.
\label{j_def}
\end{equation}
and in addition $C_0 \equiv \alpha^{[c_2,\beta,0]}$ and $a_k \equiv
\alpha^{[c_2,\beta,k]} / (e^{2k} \alpha^{[c_2,\beta,0]})$.
Comparison with equation~(\ref{def.superposition_rho}) shows that
\begin{equation}
  C_0 = q^\gamma / D^{[c_2,\beta,0]} ,
\qquad
  a_k = (\gamma/2)_k \> D^{[c_2,\beta,0]} / \> 
     (k! \> D^{[c_2,\beta,k]} ) ,
\label{def.Ca}
\end{equation}
where the factors $D^{[c_2,\beta,k]}$ are given explicitly in
equation~(\ref{def.Dfac}).

In case~II the DF is built entirely with components $f_{\rm
e}^{[c_2,\mu,k]}$ for which $\mu = \beta+k$, where $\beta$ is a
constant. The DF is then:
\begin{equation}
  f_{\rm e}^{\rm II} ({\cal E},\zeta^2,\eta^2) = 
      C_0 \> g({\cal E}) \> \zeta^{-2\beta} \> h(e^2 \eta^2 / \zeta^2) ,
\label{dfII_even}
\end{equation}
where the constant $C_0$ and the function $g({\cal E})$ are as defined for
case~I, the function $h$ is defined as
\begin{equation}
  h(e^2 \eta^2 / \zeta^2) \equiv \sum_{k=0}^{\infty} b_k \> 
                                 (e^2 \eta^2 / \zeta^2)^k ,
\label{h_def}
\end{equation}
and $b_k \equiv \alpha^{[c_2,\beta+k,k]} / (e^{2k}
\alpha^{[c_2,\beta,0]})$. Equation~(\ref{def.superposition_rho})
now gives 
\begin{equation}
  b_k \> = \> { {D^{[c_2,\beta,0]} \> \left({\gamma\over2}\right)_k } \over
                { k! \> D^{[c_2,\beta+k,k]}} }
      \> = \> { { (1)_k \left({\gamma\over2}\right)_k } \over
                { k! \> \left({{1}\over{2}}\right)_k } } .
\label{def.Cb}
\end{equation}
The second equality follows upon substitution of
equation~(\ref{def.Dfac}) and some algebraic manipulations. The
series~(\ref{h_def}) is therefore recognized as a hypergeometric
function:
\begin{equation}
  h (e^2 \eta^2 / \zeta^2) = 
     {}_2F_1 (1, \fr{\gamma}{2}; \fr12; e^2 \eta^2 / \zeta^2). 
\label{h2F1} 
\end{equation} 
Recall that the generalized hypergeometric function ${}_pF_{q}$ is
defined as
\begin{equation}
  {}_pF_{q}\left({\alpha}_{1},\ldots,{\alpha}_{p};
	         {\beta }_{1},\ldots,{\beta }_{q}; x \right) =
  \sum_{k=0}^{\infty} { {(\alpha_1)_k \cdots (\alpha_p)_k} \over
      {(\beta_1)_k \cdots (\beta_q)_k} } {{x^k}\over{k!}}
\label{def.hypergeometric} ,
\end{equation}
for $0 \leq x \leq 1$, ${\alpha}_{l} > 0$ for $l=1,\ldots,p$,
${\beta}_{m} > 0$ for $m=1,\ldots,q$. It sometimes reduces to an
elementary function for special values of the parameters. One always
has ${}_pF_{q}\,(\ldots;\ldots;0) \equiv 1$. The function $h$ in
equation~(\ref{h2F1}) is elementary for integer values of
$\gamma$. The explicit expressions for $\gamma = 1, 2, 3$ and 4 are
given in Table~1 (and are illustrated in Figure~1 below).

For both case~I and case~II, the DF is positive definite if and only if both
\begin{equation}
  \beta < 1 , 
\qquad\qquad
  {\rm and}
\qquad\qquad
  \gamma \> > \> (\delta/2) + \beta (2-\delta) .
\label{DFpos}
\end{equation}
If $\gamma > 3/2$, then the latter constraint is satisfied for all
$\beta<1$ and $0 \leq \delta \leq 1$. The DF for either case is easily
evaluated numerically using the series expansion of $j$ or $h$,
respectively. These series generally converge quickly.

\subsubsection{Special cases}

There are special cases for which the case~I and case~II DFs
simplify. Some are collected here.

In the spherical case $q=1$, one has $j = h = 1$, and $f_{\rm e}^{\rm
I}$ and $f_{\rm e}^{\rm II}$ are identical. The DFs now depend only on
${\cal E}$ and ${\zeta}^2$, and not on $L_z^2$. They describe
constant-anisotropy models (e.g., H\'enon 1973). The value of $\beta$
controls the velocity dispersion anisotropy (see Section~2.3).

The $\beta \rightarrow -\infty$ limit yields the model with all stars
on circular orbits (for which $\zeta^2 = 1$):
\begin{equation}
  \lim_{\beta \rightarrow -\infty} f_{\rm e}^{\rm I}  =
  \lim_{\beta \rightarrow -\infty} f_{\rm e}^{\rm II} = \>
     K_{-\infty} \> g({\cal E}) \> \delta_{\rm D}(1-\zeta^2) 
                 \> h(e^2 \eta^2) ,
\qquad\qquad \!\!\!\!\!\!
  K_{-\infty} \> \equiv \> {{2q^\gamma}\over{\upi^2}} \> \times \>
     \left\{ \begin{array}{ll}
        \exp(\gamma/2), & \qquad \delta = 0 ; \\
        \left ( {{2-\delta}\over2} \right )^{(\delta-\gamma)/\delta} ,
                          & \qquad 0 < \delta \leq 1 .
     \end{array}\right.
\label{df_circ}
\end{equation}
This model is physical for all relevant $\gamma$ and $\delta$,
cf.~equation~(\ref{DFpos}). The function $\delta_{\rm D}(\ldots)$ is
Dirac's delta--function. The equality of the case~I and case~II DFs in
this limit follows directly from the construction of the DFs. The
case~I DFs are built from components $f_{\rm e}^{[c_2,\mu,k]}$ with
$\mu = \beta$, the case~II DFs from components with $\mu =
\beta+k$. For $\beta \rightarrow -\infty$ these components
are~identical.

The $\beta \rightarrow 1$ limit of the case~II DF yields the model
with all stars on radial orbits (for which $\zeta^2 = 0$):
\begin{equation}
  \lim_{\beta \rightarrow 1} f_{\rm e}^{\rm II} \> = \>
     K_{1} \> g({\cal E}) \> \delta_{\rm D}(\zeta^2)
                 \> (1 - e^2 \sin^2 \theta)^{-\gamma /2} ,
\qquad\qquad
  K_{1} \> \equiv \> {{q^\gamma}\over{\upi^{3/2}}} \> \times \>
     \left\{ \begin{array}{ll}
        2 (\gamma-2)^{1/2} \exp(1), & \qquad \delta = 0 ; \\
        2 \delta^{1/2} ({2 \over {2-\delta}})^{{2-\delta}\over{\delta}}
           { {\Gamma({1\over \delta} [\gamma+\delta-2])} \over
             {\Gamma({1\over \delta} [\gamma+\delta-2] - {1\over2})} },
                          & \qquad 0 < \delta \leq 1 .
     \end{array}\right.
\label{df_rad}
\end{equation}
This model is physical only for $\gamma > 2 - \delta/2$,
cf.~equation~(\ref{DFpos}). The quantity $\sin^2 \theta$ is an
integral of motion for radial orbits, and this DF is thus indeed a
solution of the collisionless Boltzmann equation. The $\beta
\rightarrow 1$ limit of the case~I DF does {\it not\/} yield a model 
with all stars on radial orbits (see Figure~3 below).

When $\beta=0$, the case~I DF depends only on ${\cal E}$ and
${\eta}^2$, and hence is independent of $L^2$. This is the classical
axisymmetric $f(E, L_z)$ model.

When the potential is Keplerian ($\delta=1$), the function $j$ that appears in
the case~I DFs can be expressed in terms of a generalized hypergeometric
series. Equation~(\ref{def.Ca}) gives for this case
\begin{equation}
  C_0 = { {q^\gamma 4^\beta} \over
               {{\cal B}({1\over2}, {1\over2}) \>
                {\cal B}({1\over2}, 1\!-\!\beta) \>
                {\cal B}({3\over2}\!-\!\beta,
                         \gamma\!-\!\beta\!-\!{1\over2})} } ,
\qquad
  a_k = { { (1)_k \left({\gamma\over2}\right)_k 
    \left({{\gamma -2\beta +1}\over{2}}\right)_k
    \left({{\gamma -2\beta +2}\over{2}}\right)_k } \over
    { k! \> \left({{1}\over{2}}\right)_k (1\!-\!\beta)_k
    \left(\gamma\!-\!\beta\!-\!{{1}\over{2}}\right)_k } } ,
\qquad
  ({\rm for} \>\> \delta=1) .
\label{KeplerCa}
\end{equation}
The series~(\ref{j_def}) is thus
\begin{equation} 
  j (e^2 \eta^2) = 
     {}_4F_3 (1, \fr{\gamma}{2}, \fr{\gamma -2\beta +1}{2},
                   \fr{\gamma -2\beta+2}{2}; \fr12, 
                   1\!-\!\beta, \gamma\!-\!\beta\!-\!\fr12; 
                   e^2 \eta^2), \qquad
  ({\rm for} \>\> \delta=1) .
\label{Keplerj}
\end{equation}
This reduces to an elementary function when $\gamma$ and $2\beta$ are
integers. Explicit expressions for $\gamma = 1, 2, 3$ and 4 are given
in Table~1 (and are illustrated in Figure~1 below). The expressions
for $\gamma = 1$, 2 and~3 are elementary for arbitrary $\beta$. In the
special case $\beta=0$, one has (cf.~eqs. [\ref{KeplerCa}]
and~[\ref{Keplerj}]) that
\begin{equation}
  C_0 = { {q^\gamma} \over
               {2 \upi \> {\cal B}(\gamma\!-\!{1\over2},{3\over2})} } , 
\qquad
  j (e^2 \eta^2) = 
      {}_3F_2 ( \fr{\gamma}{2}, \fr{\gamma+1}{2}, \fr{\gamma+2}{2}; 
                    \fr12, \fr{2\gamma - 1}{2}; e^2 \eta^2),
\qquad
  ({\rm for} \>\> \delta=1, \> \beta=0) .
\label{Keplerjspher}
\end{equation}
This reproduces the $f(E, L_z)$ scale-free large-radii limit given in
equation~(3.24) of Qian \etal (1995) for general $\gamma$, and in
equation~(B2) of Dehnen \& Gerhard (1994) for $\gamma=4$. The
elementary expressions for integer $\gamma$ follow from those given in
Table~1 upon substitution of $\beta=0$.

\subsubsection{Odd part}

The mass density determines uniquely the part of the DF even in
$L_z$. The odd part, $f_{\rm o} ({\cal E},{\zeta}^2,{\eta})$, can be
specified freely, with the only constraint that the total DF $f =
f_{\rm e} + f_{\rm o}$ be positive definite. A natural choice is
\begin{equation}
  f_{\rm o} ({\cal E},{\zeta}^2,{\eta}) \equiv
      (2s-1) \> {\rm sign} (\eta) \> \eta^{2t} \> 
      f_{\rm e} ({\cal E},{\zeta}^2,{\eta}^2) ,
\label{df_odd}
\end{equation}
where $0 \leq s \leq 1$ and $t \geq 0$ are two free parameters. The
fraction of stars on circular orbits in the equatorial plane that
rotates clockwise is equal to $s$. A model with $s=1/2$ is
non--rotating. The parameter $t$ determines the extent to which the
net rotation of the model stems from high-angular momentum orbits.
The odd part with $t=0$ and either $s=0$ or $s=1$ is referred to as
`the maximally rotating odd part'. All stars with $L_z \not= 0$ have
the same sense of rotation around the symmetry axis in a model with
this odd part. The DF that generates the largest amount of mean
streaming consistent with the given mass density and potential is the
$\beta \rightarrow -\infty$ model with the maximally rotating odd
part. This model is referred to as `the maximum streaming model'.

\subsubsection{Density of states}

The amount of mass contributed by stars on orbits with given $({\cal
E},\zeta^2,\eta^2)$ is not determined solely by the DF, but also by
the `density of states' (e.g., Binney \& Tremaine 1987). In the
present context the density of states $w({\cal E},\zeta^2,\eta^2)$ is
defined through the following expression for the total mass of the
system:
\begin{equation}
  M = \int_{\Psi_{\infty}}^{\Psi(0)} {\rm d} {\cal E}
      \int_{0}^{1} {\rm d} \zeta^2
      \int_{0}^{\zeta^2} {\rm d} \eta^2 \>
           w({\cal E},\zeta^2,\eta^2) \> f({\cal E},\zeta^2,\eta^2) .
\label{Mass_state}
\end{equation}

To obtain an explicit expression for the density of states for the
case of an axisymmetric mass density in a spherical potential, one
expresses the total mass $M$ as the integral of $2 \upi r^2 \, \sin
\theta \, \rho(r,\theta)$ over ${\rm d}r \> {\rm d} \theta$. Then
$\rho(r,\theta)$ is substituted from equation~(\ref{rhointf2}), and
the integration variables are transformed to $({\cal
E},\zeta^2,\eta^2)$.  Rearrangement of the order of the integrations
then yields an expression for the density of states as a
two-dimensional integral over ${\rm d}r \> {\rm d} \theta$. For a
spherical potential the integral over ${\rm d} \theta$ can be
evaluated analytically, with the final result
\begin{equation}
  w({\cal E},\zeta^2,\eta^2) \> = \> 
     2 \upi^2 \> L_{\rm max} ({\cal E}) \> (\zeta^2)^{-1/2} \>
     (\eta^2)^{-1/2} \>
        \int_{r_{-}({\cal E},\zeta^2)}^{r_{+}({\cal E},\zeta^2)}
            r \> \lbrace r^2 [\Psi(r)-{\cal E}] L^{-2}_{\rm max} ({\cal E})
                  - \zeta^2 \rbrace^{-1/2} \> {\rm d}r .
\label{densstate}
\end{equation}
The radii $r_{\pm}$ are the roots of the expression in curly
braces. The interval $[r_{-},r_{+}]$ contains the radii accessible to
a star with given $({\cal E},\zeta^2)$. The integral in
equation~(\ref{densstate}) can be calculated analytically only for the
Kepler potential, $\delta=1$. In this case
\begin{equation}
  w({\cal E},\zeta^2,\eta^2) \> = \>
     {{\upi^3}\over{4}} \> {\cal E}^{-5/2} \> (\zeta^2)^{-1/2} \>
     (\eta^2)^{-1/2}  ,
  \qquad ({\rm for} \>\> \delta=1) . 
\label{densstate_kep}
\end{equation}
Equation~(\ref{Mass_state}) can be used to calculate the differential
mass distribution as function of the integrals of motion. For example,
for a spherical mass density in a Kepler potential one has, for either
the case~I or the case~II DF (cf.~Section~2.2.3), $f =
C_0 \, {\cal E}^{\gamma - (3/2)} \, \zeta^{-2\beta}$, and thus
\begin{equation}
  { {{\rm d}M} \over { {\rm d}{\cal E} \> {\rm d}\zeta^2 } } \> = \> 
       { {\upi^3}\over {2} } \> C_0 \> {\cal E}^{\gamma -4} \>
       (\zeta^{2})^{-\beta}  ,
\qquad ({\rm for} \>\> \delta=1, \> q=1) .
\end{equation}

\begin{table*}
\renewcommand{\arraystretch}{1.55}

\begin{minipage}{437truept}
\caption{The intrinsic second order velocity moments. Here $x^2 \equiv 
e^2 \sin^2\theta$. The results are elementary for case~II. For case~I
the second order velocity moments reduce to hypergeometric functions
for the Kepler potential. These reduce to elementary functions when
$\gamma$ and $2\beta$ are integers. This is illustrated by the results
for $\beta=0$ listed in the bottom part of the Table. The ${}_3F_2$
function that occurs in the expression for $\langle v_\phi^2 \rangle$
for this case can be evaluated by means of the relation ${}_3F_{2}
({{\gamma}\over{2}}, {{\gamma+1}\over{2}}, {{3}\over{2}}; {1\over2},
{{\gamma+3}\over{2}}; x^2) = {{\rm d}\over {{\rm d}x}} \lbrack x \>
{}_2F_{1} ({{\gamma}\over{2}}, {{\gamma+1}\over{2}};
{{\gamma+3}\over{2}}; x^2) \rbrack$. For $\gamma=1$ the listed
expressions for $\langle v_\theta^2 \rangle$ and $\langle v_{\phi}^2
\rangle$ are independent of $\beta$, and hence these results are valid
for all $\beta$. The quantity $\langle v_r^2 \rangle$ does depend on
$\beta$ for $\gamma=1$. Cases not covered in this Table can be
calculated with the series expansion~(42).}
\begin{tabular*}{437truept}{l}
\hline
\noalign{\smallskip}
\hbox to 425truept{\hfill
Case~II DF, arbitrary $\delta$, $\beta$ and $\gamma$
\hfill} \\
\noalign{\smallskip}
\hline
\noalign{\smallskip}
\end{tabular*}
\end {minipage}

\begin{minipage}{437truept}
\quad\quad
\begin{tabular*}{395truept}{l@{\extracolsep{\fill}} l}
$\langle v_r^2 \rangle = [2 \> (\delta + \gamma - 2\beta) \> r^{\delta}]^{-1}$
& $\langle v_{\theta}^2 \rangle = 
      (1-\beta) \langle v_r^2 \rangle \times \left\{ {\begin{array}{ll}
      -{(1-x^2) \over x^2} \ln (1-x^2) 
         &\qquad\quad \,\>\> (\gamma=2)     \\
      {2[(1-x^2)-(1-x^2)^{\gamma/2}] \over (\gamma-2) x^2}
                                        &\qquad\quad \,\>\> (\gamma\not=2) \\
                                                     \end{array}}\right.$   \\
& $\langle v_{\phi}^2 \rangle = 
      (1-\beta) \langle v_r^2 \rangle \times \left\{ {\begin{array} {ll} 
      2 + {(1-x^2) \over x^2} \ln (1-x^2) 
                                          &\qquad (\gamma=2) \\
      {2[(1-x^2)^{\gamma/2} +(\gamma-1)x^2 -1] 
                          \over (\gamma-2)x^2}
                                          &\qquad (\gamma\not=2) \\
                                                     \end{array}}\right.$ \\
\noalign{\medskip}
\end{tabular*}
\end{minipage}

\begin{minipage}{437truept}
\begin{tabular*}{437truept}{l}
\hline
\noalign{\smallskip}
\hbox to 425truept{\hfill
Case~I DF, Kepler potential ($\delta=1$), arbitrary
            $\beta$ and $\gamma$
\hfill} \\
\noalign{\smallskip}
\hline
\noalign{\smallskip}
\end{tabular*}
\end {minipage}

\begin{minipage}{437truept}
\quad\quad\quad\quad\quad\quad\quad
\begin{tabular*}{302truept}{l@{\extracolsep{\fill}} l}
& $\langle v_{r}^2 \rangle = \>\>
  {(1 - x^2)^{\gamma/2} \over 2r (\gamma - 2\beta + 1)} \>\>
        {}_2F_{1} ({{\gamma}\over{2}},{{\gamma-2\beta+1}\over{2}};
                   {{\gamma-2\beta+3}\over{2}}; x^2) $ \\
& $\langle v_{\theta}^2 \rangle = \>\>
  { (1 - x^2)^{\gamma/2} \over 2r (\gamma - 2\beta + 1)} \>\>
  (1\!-\!\beta) \>\>
        {}_4F_{3} (1,{{\gamma}\over{2}},{{\gamma -2\beta +1}\over{2}},
                   2\!-\!\beta; 1\!-\!\beta,2, 
                              {{\gamma-2\beta+3}\over{2}}; x^2)$ \\
& $\langle v_{\phi}^2 \rangle = \>\>
  {(1 - x^2)^{\gamma/2} \over 2r (\gamma - 2\beta + 1)} \>\>
  (1\!-\!\beta) \>\>
        {}_5F_{4} (1,{{\gamma}\over{2}},{{\gamma -2\beta +1}\over{2}},
                   {{3}\over{2}},2\!-\!\beta;{{1}\over{2}},1\!-\!\beta,2,
                   {{\gamma-2\beta+3}\over{2}};x^2)$  \\
\noalign{\medskip}
\end{tabular*}
\end{minipage}

\begin{minipage}{437truept}
\begin{tabular*}{437truept}{l}
\hline
\noalign{\smallskip}
\hbox to 425truept{\hfill
Case~I DF, Kepler potential ($\delta=1$), $f(E,L_z)$ model
            ($\beta=0$), and integer $\gamma$
\hfill} \\
\noalign{\smallskip}
\hline
\noalign{\smallskip}
\end{tabular*}
\end {minipage}

\begin{minipage}{437truept}
\begin{tabular*}{437truept}{
  l@{\hspace{1.5truecm}} l@{\hspace{1.5truecm}} l}
$\gamma$  & $\langle v_{r}^2 \rangle = \langle v_\theta^2 \rangle
            = {(1-x^2)^{\gamma/2} \over 2r(\gamma+1)} \>
              {}_2F_{1} ({{\gamma}\over{2}}, {{\gamma+1}\over{2}}; 
                                             {{\gamma+3}\over{2}}; x^2)$ 
          & $\langle v_{\phi}^2 \rangle = 
           {(1 - x^2)^{\gamma/2} \over 2r (\gamma + 1)} 
        {}_3F_{2} ({{\gamma}\over{2}}, {{\gamma+1}\over{2}}, {{3}\over{2}};
                   {{1}\over{2}}, {{\gamma+3}\over{2}}; x^2)$  \\
\noalign{\bigskip}
1  & ${(1-x^2)^{1/2} -(1-x^2) \over 2rx^2}$ 
   & ${1-(1-x^2)^{1/2} \over 2rx^2}$ \\
2  & ${(1-x^2) \over 2rx^2} 
      \bigl( -1 + {1\over 2x} \ln {1+x \over 1-x} \bigr)$
   & ${1 \over 2rx^2} 
      \bigl( 2-x^2 -{1-x^2 \over x} \ln {1+x \over 1-x} \bigr)$ \\
3  & ${ {2 - 3 x^2 + x^4 - 2 (1-x^2)^{3/2}} \over {2rx^4} }$
   & ${ {-6 + 9 x^2 - 2 x^4 + 6 (1-x^2)^{3/2}} \over {2rx^4} }$ \\
4  & ${(1-x^2) \over 4rx^4} \bigl( 3 - 2x^2 -{3(1-x^2) \over 2x}
      \ln {1+x \over 1-x} \bigr)$ 
   & ${(1-x^2) \over 2rx^4} \bigl( 3x^2 + {{x^2} \over {1 - x^2}} - 6
      +{3(1-x^2) \over x} \ln{1+x \over 1-x} \bigr)$ \\ 
\noalign{\smallskip}
\hline 
\end{tabular*}
\end{minipage}

\renewcommand{\arraystretch}{1.0}
\end{table*}

\subsection{Intrinsic velocity moments}

The intrinsic velocity moments $\langle v_r^l \> v_{\theta}^m \>
v_{\phi}^n \rangle$ of arbitrary order follow from
\begin{equation}
  \rho \langle v_r^l \> v_{\theta}^m \> v_{\phi}^n \rangle \equiv 
     \int {\rm d}^3{\bmath v} \> f \> v_r^l \> v_{\theta}^m \> v_{\phi}^n ,
\label{def.velmoments}
\end{equation}
where $l,m,n \geq 0$ are integers. The quantities $\rho \langle v_r^l
\> v_{\theta}^m \> v_{\phi}^n \rangle$ with $l+m+n=2$ are often called
the stresses. As before, we transform to the integration variables
$(\overline{{\cal E}},\overline{{\zeta}^2}, \overline{{\eta}^2})$
(eq.~[\ref{corvelnew}]), and use the relation
\begin{equation}
  v_{r}^l \> v_{\theta}^m \> v_{\phi}^n = 
     \Psi^{(l+m+n)/2} \> (1-\overline{{\cal E}})^{(l+m+n)/2} \>
     (1-\overline{{\zeta}^2})^{l/2} \> (\overline{{\zeta}^2})^{(m+n)/2} \>
     (1-\overline{{\eta}^2})^{m/2} \> (\overline{{\eta}^2})^{n/2} ,
\end{equation}
which follows from equation~(\ref{corvelsph}). For a DF component
$f^{[c_2,\mu,\lambda]}$ as given in equation~(\ref{def.DF_comp}), with
$c_2 = \gamma - (3\delta/2)$ as before, the integral yields\newline
\vbox{
\begin{eqnarray}
\lefteqn{
  \rho \langle v_r^l \> v_{\theta}^m \> v_{\phi}^n 
           \rangle^{[c_2,\mu,\lambda]} =
    r^{-\gamma-{\delta\over2}(l+m+n)} \> \sin^{2\lambda} \theta \> 
    {\cal B} (\fr{m+1}{2}, \lambda\!+\!\fr{n+1}{2}) \>
    {\cal B} (\fr{l+1}{2}, 1\!-\!\mu\!+\!\lambda\!+\!\fr{m+n}{2}) } 
                                                      \nonumber \\ 
  & & \qquad\qquad \times \> \left\{ \begin{array}{ll}
    2^{\lambda-\mu} \> \exp(\lambda\!-\!\mu) \>
    \Gamma ({3\over2}\!-\!\mu\!+\!\lambda\!+\!{{l+m+n}\over{2}}) \>
    (\gamma\!-\!2\mu\!+\!2\lambda)^{-{3\over2}+\mu-\lambda-{{l+m+n}\over2}} ,
 & \qquad \delta = 0 ; \\
    \delta^{-(3+l+m+n)/2} \>
    \left[({\delta\over2}) ({{2-\delta}\over 2})^{{2-\delta}\over{\delta}}
         \right]^{\mu-\lambda} \>
    {\cal B} ({3\over2}\!-\!\mu\!+\!\lambda\!+\!{{l+m+n}\over{2}},
         {1\over\delta} 
             [\gamma\!-\!(2\!-\!\delta)(\mu\!-\!\lambda)]\!-\!{1\over2} ) ,
  & \qquad 0 < \delta \leq 1 , \\
       \end{array} \right.
\end{eqnarray}
}
so that the contributions of each DF component to the velocity moments 
are all elementary.

The moments $\rho \langle v_r^l \> v_{\theta}^m \> v_{\phi}^n
\rangle$ of the DFs with the case~I and case~II even parts defined in 
Section~2.2.3, and the odd part defined in equation~(\ref{df_odd}),
can be expressed as series of these components:
\begin{equation}
  \rho \langle v_r^l \> v_{\theta}^m \> v_{\phi}^n \rangle^{\rm I} =
    (2S-1) \> C_0 \> \sum_{k=0}^{\infty} a_k \> e^{2k} \>
    \rho \langle v_r^l \> v_{\theta}^m \> v_{\phi}^n 
         \rangle^{[c_2,\beta,\lambda]} ,
\qquad
  \rho \langle v_r^l \> v_{\theta}^m \> v_{\phi}^n \rangle^{\rm II} =
    (2S-1) \> C_0 \> \sum_{k=0}^{\infty} b_k \> e^{2k} \>
    \rho \langle v_r^l \> v_{\theta}^m \> v_{\phi}^n
         \rangle^{[c_2,\beta+k,\lambda]}
\label{velmoments} ,
\end{equation}
where one should substitute: $S=1$ and $\lambda=k$ for even $n$; and
$S=s$ and $\lambda = k+t$ for odd $n$. The summations are power series
in $e^2 \sin^2 \theta$. The velocity moments are easily evaluated
numerically from these power series, which generally converge quickly.
Substitution of the definitions of $a_k$ and $b_k$ in
equation~(\ref{velmoments}) shows that the velocity
moments of the case~I and case~II DFs are always identical on the
symmetry axis.

For the case~II DFs the power series in $e^2 \sin^2 \theta$ reduces as
before to a generalized hypergeometric function:\newline
\vbox{
\begin{eqnarray}
\lefteqn{\rho \langle v_r^l \> v_{\theta}^m \> 
                      v_{\phi}^n \rangle^{\rm II} \> = \>
   (2S\!-\!1) \> r^{-\gamma-{\delta\over2}(l+m+n)} \>
        \sin^{2T}\theta \> q^{\gamma} \>  
           {{{\cal B} ({{m+1}\over{2}},T\!+\!{{n+1}\over{2}})} \over
            {{\cal B} ({{1  }\over{2}},  {{  1}\over{2}})}}  
           {{{\cal B} ({{l+1}\over{2}}, 
                       1\!-\!\beta\!+\!T\!+\!{{m+n}\over{2}})} \over
            {{\cal B} ({{1}\over{2}},1\!-\!\beta)}} } \nonumber \\
 && \qquad\qquad\qquad\qquad \times \> 
    {}_3F_2(1, \fr{\gamma}{2}, T\!+\!\fr{n+1}{2};
            \fr12, 1\!+\!T\!+\!\fr{m+n}{2}; e^2 \sin^2\theta ) \nonumber \\
 && \qquad\qquad\qquad\qquad \times \left\{ \begin{array}{ll}
    2^T \> \exp(T) \>
       { {\Gamma({{3}\over{2}}-\beta+T+{{l+m+n}\over{2}})} \over
         {\Gamma({{3}\over{2}}-\beta)} } 
       { {(\gamma-2\beta+2T)^{-{{3}\over{2}}+\beta-T-{{l+m+n}\over{2}}} } \over
         {(\gamma-2\beta)^{-{{3}\over{2}}+\beta}} } 
  , & \qquad \delta = 0 ; \\
       \delta^{-(l+m+n)/2} \>
       \left[ ({\delta\over2}) ({{2-\delta}\over 2})^{{2-\delta}\over{\delta}}
         \right]^{-T} \>
       { { {\cal B} ({{3}\over{2}}-\beta+T+{{l+m+n}\over{2}},
                     {1\over\delta}[\gamma-(2-\delta)(\beta-T)]
                     -{1\over2} ) } \over
         { {\cal B} ({{3}\over{2}}-\beta,
                     {1\over\delta}[\gamma-(2-\delta)\beta]
                     -{1\over2} ) } }
  , & \qquad 0 < \delta \leq 1 .
                   \end{array}\right.
\label{caseIImoment}
\end{eqnarray}
}
For the case~I DFs the power series in $e^2 \sin^2 \theta$ reduces to
a generalized hypergeometric function only for the Kepler potential
($\delta=1$):\newline
\vbox{
\begin{eqnarray}
\lefteqn{\rho \langle v_r^l \> v_{\theta}^m \> 
                      v_{\phi}^n \rangle^{\rm I} = } \nonumber \\
  && (2S\!-\!1) \> r^{-\gamma-{{l+m+n}\over{2}}} \>
        (4\sin^2\theta)^T \> q^{\gamma} \>  
           {{{\cal B} ({{m+1}\over{2}},T\!+\!{{n+1}\over{2}})} \over
            {{\cal B} ({{1  }\over{2}},  {{  1}\over{2}})}}  
           {{{\cal B} ({{l+1}\over{2}}, 
                        1\!-\!\beta\!+\!T+\!{{m+n}\over{2}})} \over
            {{\cal B} ({{1}\over{2}},1\!-\!\beta)}}  
           {{{\cal B} ({{3}\over{2}}\!-\!\beta\!+\!T+\!{{l+m+n}\over{2}},
                           \gamma\!-\!\beta\!-\!{{1}\over{2}}\!+\!T)} \over
            {{\cal B} ({{3}\over{2}}-\beta,\gamma-\beta-{{1}\over{2}})}}
        \nonumber \\
 && \qquad \qquad \qquad \times \> {}_7F_6( 
       1, \fr{\gamma}{2}, \fr{\gamma-2\beta+1}{2},
        \fr{\gamma-2\beta+2}{2}, \gamma\!-\!\beta\!-\!\fr12\!+\!T,
        T\!+\!\fr{n+1}{2}, 1\!-\!\beta\!+\!T\!+\!\fr{m+n}{2}; 
         \fr12, 1\!-\!\beta, 
        \gamma\!-\!\beta\!-\!\fr12, \nonumber \\
 && \qquad\qquad \qquad \qquad\qquad 
        1\!+\!T\!+\!\fr{m+n}{2}, 
        \fr{\gamma-2\beta+1}{2}\!+\!T\!+\!\fr{l+m+n}{4},
        \fr{\gamma-2\beta+2}{2}\!+\!T\!+\!\fr{l+m+n}{4};
        e^2 \sin^2\theta ) , \qquad ({\rm for} \>\> \delta=1) .
\end{eqnarray}
}
In these equations one should substitute: $S=1$ and $T=0$ for even $n$; and
$S=s$ and $T=t$ for odd $n$. 

For the low-order velocity moments the ${}_pF_q$ in these expressions
often simplify. This is illustrated by Table~2, which lists second
order velocity moments for the case~I and case~II DFs. For case~II
these are elementary for {\it all\/} $\delta$, $\gamma$ and
$\beta$. For case~I the second order velocity moments in a Kepler
potential are elementary when $\gamma$ and $2\beta$ are integers.

It is useful to consider Binney's (1980) velocity dispersion
anisotropy function $\beta_{\rm B}$, defined as
\begin{equation}
  \beta_{\rm B} (r,\theta) \equiv 1 - {{\langle v_{\theta}^2 \rangle + \langle
    v_{\phi}^2 \rangle} \over {2 \langle v_{r}^2 \rangle}} .
\label{def.beta}
\end{equation}
On the minor axis ($\sin \theta=0$), and in the spherical limit
($e=0$), one has
\begin{equation}
  \langle v_{\theta}^2 \rangle = \langle v_{\phi}^2 \rangle,
\qquad
  \beta_{\rm B}(r,\theta) = \beta ,
\qquad 
  ({\rm if} \> e^2 \sin^2 \theta = 0) ,
\label{def.betaminor}
\end{equation}
for both case~I and case~II. For the special choice of case~I and
$\beta=0$ (i.e., the $f(E,L_z)$ models) one has $\langle v_{\theta}^2
\rangle = \langle v_r^2 \rangle$ everywhere, for both spherical and
flattened models. More interestingly, it follows from the expressions
in Table~2 that for case~II, Binney's anisotropy function is
independent of the polar angle:
\begin{equation}
  \beta_{\rm B}(r,\theta) = \beta 
\qquad
  ({\rm for \>\> case \>\> II \>\> and \>\> any \>\>} e, \> \theta).
\label{def.betaII}
\end{equation}
This means that the case~II DFs describe {\it flattened
constant-anisotropy\/} models. Models with $\beta=0$ are isotropic,
those with $\beta<0$ are tangentially anisotropic, and those with
$\beta>0$ are radially anisotropic.

\subsection{Projected velocity moments}

Following Evans \& de Zeeuw (1994), Cartesian coordinates $(x',y',z')$
are defined such that $x'$ lies along the projected major axis of the
model, $y'$ lies along the projected minor axis, and $z'$ lies along
the line of sight. The inclination of the galaxy is denoted by $i$,
such that for $i=90^{\circ}$ the object is edge--on, while for
$i=0^{\circ}$ it is face--on.

The projected mass density $\Sigma_{\rm p} (x',y')$ on the plane of
the sky can be calculated as in Fillmore (1986), which yields
\begin{equation}
  \Sigma_{\rm p} \equiv
     \int_{-\infty}^{\infty} {\rm d} z' \> \rho(r,\theta) \> = \>
     {{q}\over{q_{\rm p}}} {\cal B} (\fr{\gamma-1}{2}, \fr12) 
       \left({x'}^2 + {{{y'}^2}\over {q_{\rm p}^2}} 
       \right)^{{{-\gamma+1}\over{2}}}
\label{def.Sigma} .
\end{equation}
The quantity $\Sigma_{\rm p}$ is proportional to the surface
brightness, if the mass-to-light ratio is constant. The axial ratio of
the similar concentric elliptical isophotes is $q_{\rm p}$, which
satisfies $q_{\rm p}^2 = \cos^2 i + q^2 \sin^2 i$. Their ellipticity
is $\epsilon_{\rm p} \equiv 1 - q_{\rm p}$.

The line-of-sight velocity at any given point is $v_{z'} = C_r v_r +
C_{\theta} v_{\theta} + C_{\phi} v_{\phi}$, where
\begin{equation}
  C_r        = \sin\theta \cos\phi \sin i + \cos\theta \cos i , \qquad
  C_{\theta} = \cos\theta \cos\phi \sin i - \sin\theta \cos i , \qquad
  C_{\phi}   = -\sin\phi \sin i .
\end{equation}
The $n$-th line-of-sight velocity moment $\langle v_{z'}^n \rangle$ at
any given point satisfies
\begin{equation}
  \rho \langle v_{z'}^n \rangle = 
     \sum_{j=0}^{n} \sum_{k=0}^{n-j} {n \choose j} {{n-j} \choose k}
      C_r^j \> C_{\theta}^k \> C_{\phi}^{n-j-k} \> 
      \rho \langle v_{r}^{j} v_{\theta}^{k} v_{\phi}^{n-j-k} \rangle ,
\end{equation}
which is obtained by using the binomial theorem twice. The quantities
$\rho \langle v_{r}^{j} v_{\theta}^{k} v_{\phi}^{n-j-k} \rangle$ are
as given in equation~(\ref{velmoments}). The $n$-th
projected line-of-sight velocity moment $\langle v_{z'}^n
\rangle_{\rm p}$ on the $(x',y')$ plane of the sky follows upon
integration along the line of sight:
\begin{equation}
  \langle v_{z'}^n \rangle_{\rm p} \equiv
    {1\over{\Sigma_{\rm p}}} \> \int_{-\infty}^{+\infty}
    {\rm d} z' \> \rho \langle v_{z'}^n \rangle .
\label{losmom}
\end{equation}
This integral must generally be evaluated numerically.

\subsection{Velocity profiles}

The velocity profile (VP) at any point $(x',y')$ on the sky is the 
line-of-sight velocity distribution of the stars:
\begin{equation}
  {\rm VP} (v_{z'}) \equiv
    {1 \over {\Sigma_{\rm p}}}
    \int {\rm d} z' \> \int {\rm d} v_{x'} \> \int {\rm d} v_{y'} \> f ,
\label{VPdef}
\end{equation}
where $f$ is the DF. The integration limits are set by the fact that
the integrand $f$ is zero for those points in phase space where ${\cal
E} > \Psi$. The moments of the VP are equal to the projected
line-of-sight velocity moments given in equation~(\ref{losmom}), i.e.,
\begin{equation}
  \int_{-\infty}^{\infty} {\rm d} v_{z'} \> [{\rm VP} (v_{z'}) ]^n
  = \langle v_{z'}^n \rangle_{\rm p} .
\end{equation}
Observed VPs are often represented as a Gauss--Hermite series (e.g., van der
Marel \& Franx 1993; Gerhard 1993). This series is parametrized by the
normalization $\gamma_{\rm G}$, mean $V$ and dispersion $\sigma$ of the
best-fitting Gaussian to the VP, and the Gauss--Hermite moments $h_3, h_4,
\ldots$, which quantify deviations of the VP from this Gaussian.  Calculation
of these quantities for the models discussed here requires knowledge
of the VP. The VP can in principle be calculated by direct numerical
evaluation of the triple integral (\ref{VPdef}), but this is
cpu--intensive and not convenient in practice. An alternative is to
recover the VP from its moments $\langle v_{z'}^n \rangle_{\rm p}$,
which can be obtained by single integrations
(cf.~eq.~[\ref{losmom}]). This is a well-known, ill-conditioned
mathematical problem, but after some experimenting an algorithm was
found that is satisfactory for the purpose at hand. It is described in
Appendix~A. It works well, except for the small region of parameter
space describing strongly flattened models with a logarithmic
potential and large anisotropy, which will not be discussed in the
remainder of the paper.

\begin{figure*}
\centerline{\epsfbox{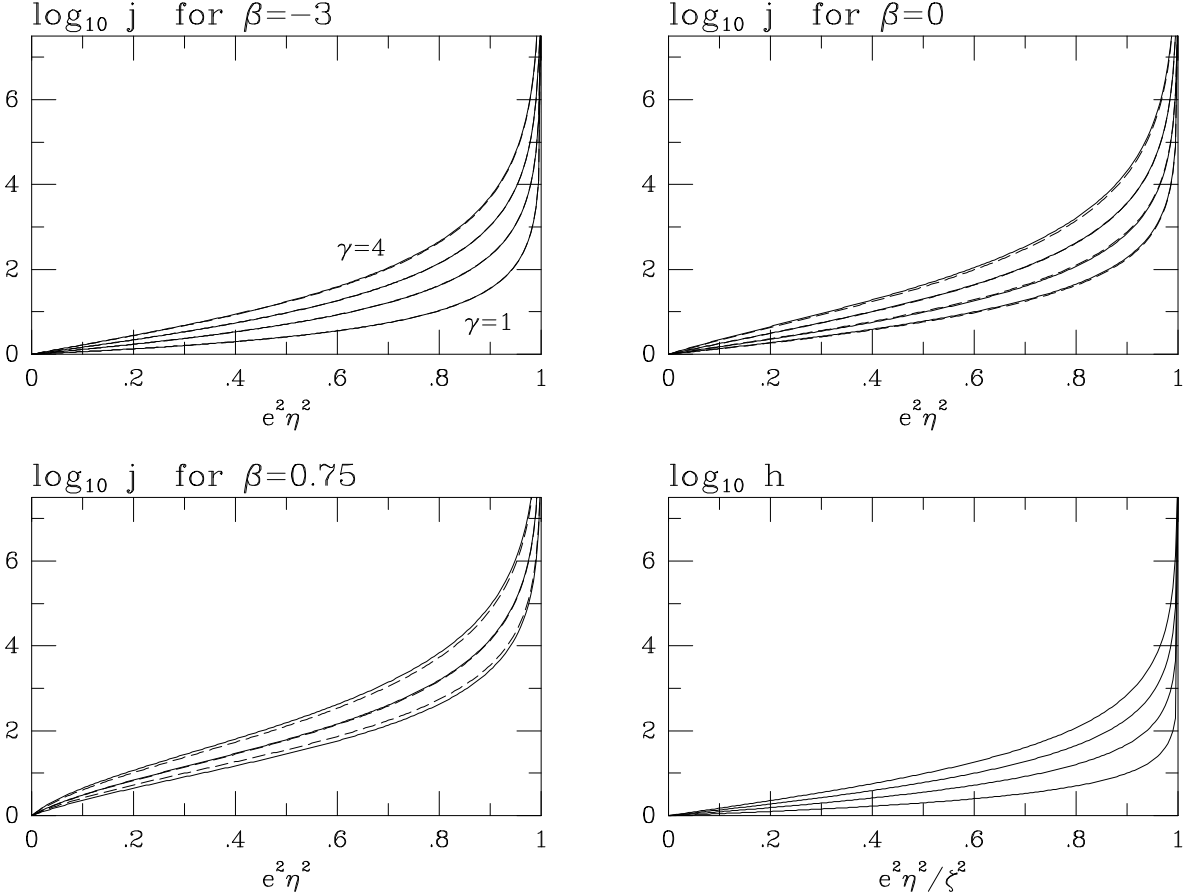}}
\caption{The functions $j(e^2 \eta^2)$ and $h(e^2 \eta^2 / \zeta^2)$
responsible for the flattening of the mass density for the case~I and
case~II DFs, respectively. Solid curves are for the Kepler potential
($\delta=1$), dashed curves for the logarithmic potential
($\delta=0$). Each panel has four curves for each choice of potential,
corresponding to mass density power-law slopes $\gamma = 1$, $2$, $3$
and $4$. The function $j$ depends on $\beta$. It is shown for three
representative values: $\beta = -3$, $0$ and $0.75$. The $\gamma=1$
curves are absent in the $\beta=0.75$ panel, because these do not
correspond to a physical DF. The function $h$ is independent of both
$\beta$ and $\delta$. The vertical scale in all panels is logarithmic.}
\end{figure*}

\section{Model properties}

\subsection{Distribution functions}

In Section~2 two families of DFs were presented (referred to as case~I
and case~II) which generate a scale-free spheroidal mass density with
power-law slope $\gamma \geq 0$ and flattening $q \leq 1$, in a
scale-free spherical potential with power-law slope $0 \leq \delta
\leq 1$. The part of the DF even in $L_z$ has (for each family) one
free parameter $-\infty < \beta < 1$, which regulates the dynamical
structure of the model. The DFs are given by
equations~(\ref{dfI_even}) and~(\ref{dfII_even}), respectively. They
are physical if and only if equation~(\ref{DFpos}) is satisfied. A
convenient ad hoc choice for the odd part of the DF is given by
equation~(\ref{df_odd}). This odd part has additional free parameters
$s$ and $t$, which regulate the mean azimuthal streaming in the model.

The DFs $f_{\rm e}^{\rm I}$ and $f_{\rm e}^{\rm II}$ have a factor
$C_0 \zeta^{-2\beta} g({\cal E})$ in common. The normalization
constant $C_0$ depends on $\gamma$, $\delta$, $\beta$ and $q$. The
quantity $\zeta^2$ is defined as the ratio $L^2 / L_{\rm max}^2({\cal
E})$ (cf.~eq.~[\ref{def.zeta_eta}]). The function $g({\cal E})$ is a
scale-free function of the energy ${\cal E}$, as required by the
nature of the density and potential. It is fully determined by
$\gamma$ and $\delta$.
 
In the spherical case one has $f_{\rm e}^{\rm I} = f_{\rm e}^{\rm II}
= C_0 \zeta^{-2\beta} g({\cal E})$. For flattened models the case~I
DFs have an extra factor $j(e^2 \eta^2)$, while the case~II DFs have
an extra factor $h(e^2 \eta^2 / \zeta^2)$. The quantity $\eta^2$ is
defined as $L_z^2 / L_{\rm max}^2({\cal E})$
(cf.~eq.~[\ref{def.zeta_eta}]). Note that $\eta^2 / \zeta^2 = L_z^2 /
L^2$. The functions $j$ and $h$, respectively, are responsible for the
flattening of the mass density. The axial ratio enters into these
functions only through the eccentricity $e$ in the argument. The
function $j$ (case~I) depends on $\gamma$, $\beta$ and $\delta$. The
function $h$ (case~II) depends only on $\gamma$. 

Figure~1 displays the functions $j$ and $h$ for $\delta=0$ and $1$,
and $\gamma = 1$, 2, 3 and 4. The function $j$ is shown for three
representative values of $\beta$: $-3$, $0$ and $0.75$. Both $j$ and
$h$ increase monotonically as function of their argument. At fixed
flattening, the physical range of the argument runs from 0 to $e^2 =
1-q^2$. For realistic elliptical galaxy models ($q \ga 0.3$), the
functions $j$ and $h$ can vary by as much as two to three orders of
magnitude over their physical range. However, this does not imply that
most of the stars in the system are on orbits with either $\eta^2=1$
(case~I) or $\eta^2 = \zeta^2$ (case~II), respectively, because the
density of states for these orbits is low (cf.~eq.~[\ref{densstate}]).

\begin{figure*}
\centerline{\epsfbox{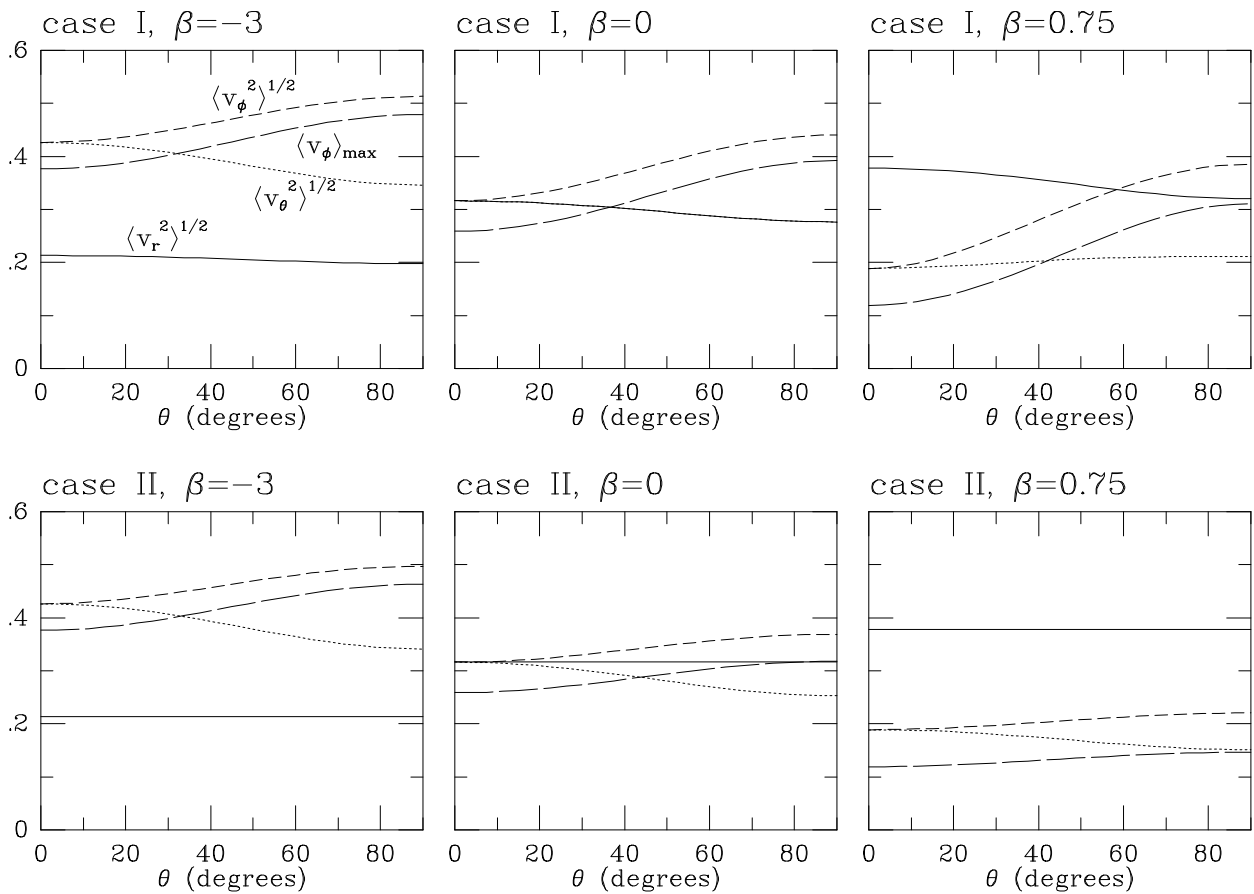}}
\caption{Velocity moments for a model with a Kepler potential
($\delta=1$) and a mass density with flattening $q=0.8$ and power-law
slope $\gamma=4$. Shown are: $\langle v_r^2 \rangle^{1/2}$ (solid
curves), $\langle v_{\theta}^2 \rangle^{1/2}$ (dotted curves),
$\langle v_{\phi}^2 \rangle^{1/2}$ (short-dashed curves) and $\langle
v_{\phi} \rangle_{\rm max}$ (long-dashed curves) as function of polar
angle $\theta$ at radius $r=1$. Top panels are for the case~I DFs,
bottom panels are for the case~II DFs. From left to right the value of
the model parameter $\beta$ is: $-3$, $0$ and $0.75$, respectively.
In the top middle panel, when $f=f(E,L_z)$, one has $\langle v_r^2
\rangle = \langle v_{\theta}^2 \rangle$. The symmetry axis has 
$\theta=0^{\circ}$, the equatorial plane has $\theta = 90^{\circ}$.}
\end{figure*}

\subsection{Intrinsic velocity moments}

To understand the dynamical structure of the models it is useful to
focus on the first and second intrinsic velocity moments. These are
easily calculated for any combination of model parameters using the
formulae in Section~2.3. As an example consider the particular case
$\gamma=4$, $q=0.8$ and $\delta=1$. Figure~2 shows for the case~I and
case~II DFs, for $\beta = -3$, $0$ and $0.75$, the dependence of
$\langle v_r^2 \rangle^{1/2}$, $\langle v_{\theta}^2
\rangle^{1/2}$ and $\langle v_{\phi}^2 \rangle^{1/2}$ on the polar
angle $\theta$, at radius $r=1$. The dependence on $r$ is simple
(see~Table~2), because of the scale-free nature of the models. The
mean azimuthal velocity $\langle v_{\phi}
\rangle_{\rm max}$ for the maximally rotating model associated with this even
part is also shown in the figure.

The case~II DFs have a constant ratio of rms radial to rms tangential
($v_{\rm t}^2 \equiv v_{\theta}^2 + v_{\phi}^2$) motion as function of
$\theta$. This ratio is determined by the model parameter $\beta$,
cf.~equation~(\ref{def.betaII}). The models with $\beta
\rightarrow 1$ have only radial orbits (with $L^2 = 0$), while the
models with $\beta \rightarrow -\infty$ have only tangential orbits
(with $L^2 = L_{\rm max}^2({\cal E})$). The quantity $\langle v_r^2
\rangle$ is constant as function of $\theta$ for the case~II DFs, 
and hence so is $\langle v_{\theta}^2 + v_{\phi}^2 \rangle$. What does
vary as function of $\theta$ is the ratio $\langle v_{\phi}^2
\rangle / \langle v_{\theta}^2 \rangle$. It is unity on the minor
axis, and increases monotonically with $\theta$.

On the symmetry axis the velocity moments for the case~I DFs are
identical to those for the case~II DFs. Away from the symmetry axis
they behave differently. The ratio $\langle v_{\phi}^2
\rangle / \langle v_{\theta}^2 \rangle$ for case~I increases 
monotonically with $\theta$, as for case~II. By contrast, the ratio
$\langle v_{\rm t}^2 \rangle / \langle v_r^2 \rangle$ also increases
monotonically with $\theta$, rather than being constant, as for
case~II. This is not very pronounced in the top left panel of Figure~2
for $\beta=-3$, because the case~I and case~II DFs are identical in
the limit $\beta \rightarrow -\infty$. However, it is very clear in
the top right panel for $\beta=0.75$. In fact, the case~I $\beta=0.75$
model has $\langle v_t^2 \rangle < \langle v_r^2 \rangle$ on the
symmetry axis, and $\langle v_t^2 \rangle > \langle v_r^2
\rangle$ in the equatorial plane.

The dependence of the intrinsic first and second order velocity
moments on $\beta$ is further illustrated in Figure~3. This figure
shows ratios of various velocity moments in the equatorial plane and
on the symmetry axis, as function of $\beta / (2-\beta)$ (this choice
of abscissa is useful because it maps the infinite interval $-\infty <
\beta < 1$ to the finite interval $-1 <
\beta / (2-\beta) < 1$). The figure clearly demonstrates the equality 
of the case~I and case~II DFs for $\beta \rightarrow -\infty$. It also
shows that the case~I DFs are always more tangentially anisotropic in
the equatorial plane than the case~II DFs, while on the symmetry axis
they are identical. The ratio $\langle v_{\phi}
\rangle_{\rm max} / \langle v_{r}^2 + v_{\phi}^2 \rangle^{1/2}$
of the mean azimuthal streaming and the rms motion in the equatorial
plane ($\theta=\upi/2$) is a monotonically decreasing function of
$\beta$. The maximum possible relative importance of mean streaming
thus decreases as the importance of radial pressure in supporting the
shape of the system increases.

\begin{figure*}
\centerline{\epsfbox{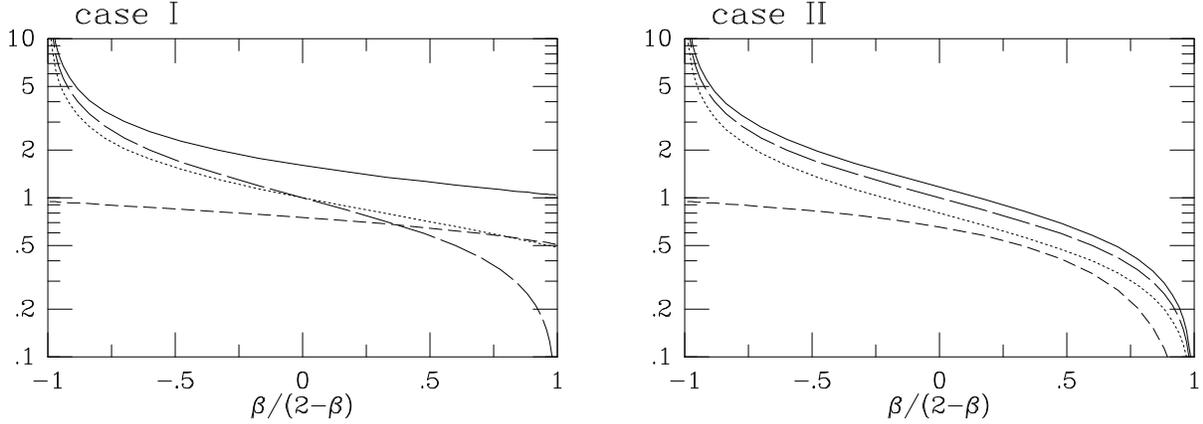}}
\caption{Ratios of various velocity moments as function of 
$\beta / (2-\beta)$, for a model with a Kepler potential
($\delta=1$) and a mass density with flattening $q=0.8$ and power-law
slope $\gamma=4$. The left edge of each panel corresponds to $\beta
\rightarrow -\infty$, the right edge to $\beta \rightarrow 1$. Shown
are the values in the equatorial plane of $[ \langle v_{\phi}^2
\rangle / \langle v_r^2 \rangle ]^{1/2}$ (solid curves), $[
\langle v_{\theta}^2 \rangle / \langle v_r^2 \rangle ]^{1/2}$ (dotted
curves), $\langle v_{\phi} \rangle_{\rm max} / \langle v_r^2 +
v_{\phi}^2 \rangle^{1/2}$ (short-dashed curves), and also the value of
$[ \langle v_{\phi}^2 \rangle / \langle v_r^2 \rangle ]^{1/2} = [
\langle v_{\theta}^2 \rangle / \langle v_r^2 \rangle ]^{1/2}$ on the
symmetry axis (long-dashed curves). The case~I DFs are always more
tangentially anisotropic in the equatorial plane than the case~II DFs,
while on the symmetry axis they are identical.}
\end{figure*}

\subsection{Projected velocity moments}

A useful observational indicator of the dynamical structure of a
stellar system is the ratio $\nu$ of the rms projected line-of-sight
velocity on the major and minor axes (van der Marel 1991):
\begin{equation}
  \nu \> \equiv \> \langle v_{z'}^2 \rangle_{\rm p,maj} \, / \,
                   \langle v_{z'}^2 \rangle_{\rm p,min}
      \> = \> (\sigma_{\rm p}^2 + v_{\rm p}^2)_{\rm maj} \, / \,
              (\sigma_{\rm p}^2)_{\rm min}  ,
\label{nudef}
\end{equation}
where $\langle v_{z'}^2 \rangle_{\rm p}$ is defined in
equation~(\ref{losmom}), and $v_{\rm p}$ and $\sigma_{\rm p}$ are the
observed mean streaming and dispersion. This ratio depends only on the
even part of the DF. It is generally a function of radius. However, in
our scale-free models it has a constant value, which can be evaluated
numerically as described in Section~2.4.

Figure~4 shows $\nu$ as function of $\beta/(2-\beta)$ for the case~I
and case~II DFs, for a system with $\gamma=4$, $\delta=1$ and
projected axial ratio $q_{\rm p} = 0.8$. The different curves correspond to
different values of the intrinsic axial ratio $q$, and hence to
different inclination angles. For the case~I DFs $\nu$ is generally an
increasing function of $\beta$, although $\nu$ is close to constant
for $-\infty < \beta \la 0$. The flatter case~I models with smaller
inclination angles have larger $\nu$. For the case~II DFs $\nu$ is a
decreasing function of $\beta$, the more steeply so for the flatter
models with smaller inclination angles. The results in Figure~4 are
generic for other values of $\gamma$, $\delta$ and $q_{\rm p}$.

Following Binney, Davies \& Illingworth (1990), van der Marel (1991)
used solutions of the Jeans equations to compare the predictions of
models with $f=f(E,L_z)$ to kinematical data for 37 elliptical
galaxies. He concluded that these models generally predict values of
$\nu$ that are too large compared to the observed values, the more so
for smaller inclinations. For the models discussed here, $f=f(E,L_z)$
corresponds to case~I with $\beta=0$. Figure~4 shows that none of the
case~I models can produce values of $\nu$ that are appreciably smaller
than those of the edge-on $f(E,L_z)$ models. From this it follows that
the case~I DFs (or their self-consistent generalizations) are probably
not a good representation of real elliptical galaxies. This can be
attributed to their property that the ratio of rms tangential to rms
radial motion always increases strongly with $\theta$. Apparently,
this is not realized in nature, although it does lead to dynamically
acceptable models. The galaxies in the van der Marel (1991) sample
have values of $\nu$ roughly between $0.9$ and $1.3$. The case~II DFs
in Figure~4 can easily reproduce this range. If elliptical galaxies
have DFs similar to that of the case~II models, Figure~4 indicates
that they will most likely have $\beta \ga 0$. This is consistent with
expectation based on N-body simulations of galaxy formation through
dissipationless collapse (van Albada 1982; Bertin \& Stiavelli 1993).

\begin{figure*}
\centerline{\epsfbox{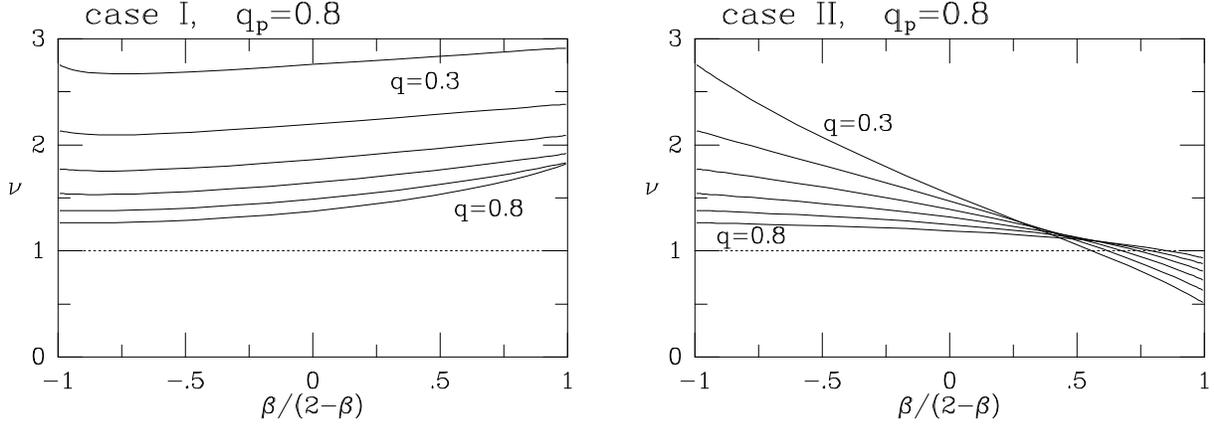}}
\caption{The ratio $\nu$ of the rms projected line-of-sight velocity 
on the major and minor axes, as function of $\beta / (2-\beta)$, for
a model with a Kepler potential ($\delta=1$) and a mass density
with power-law slope $\gamma=4$ and projected axial ratio
$q_{\rm p}=0.8$. The left edge of each panel corresponds to $\beta
\rightarrow -\infty$, the right edge to $\beta
\rightarrow 1$. The curves correspond to values of the intrinsic axial 
ratio $q = 0.3$, $0.4$, $0.5$, $0.6$, $0.7$ and $0.8$, each
corresponding to a model viewed at a different inclination angle. The
dotted lines correspond to $\nu=1$. In real galaxies $\nu$ is
generally between $0.9$ and $1.3$, indicating they are probably best
fit by case~II DFs with $\beta > 0$.}
\end{figure*}

Figure~5 shows the ratio $v_{\rm p}/\sigma_{\rm p}$ of the mean
streaming and velocity dispersion on the projected major axis for the
same set of models as in Figure~4, for the maximally rotating odd
part. Models with lower inclination and smaller axial ratio $q$
generally have larger $v_{\rm p}/\sigma_{\rm p}$, in spite of the fact
that they have less of their intrinsic streaming along the line of
sight. For both DF families $v_{\rm p}/\sigma_{\rm p}$ decreases with
$\beta$. In models with more radial motion one thus expects to see
relatively less streaming.  Bright elliptical galaxies generally have
$v_{\rm p}/\sigma_{\rm p} \la 0.4$. Figure~5 therefore shows that
bright elliptical galaxies rotate much slower than allowed
dynamically, as is well-known (e.g., Binney 1976).

\begin{figure*}
\centerline{\epsfbox{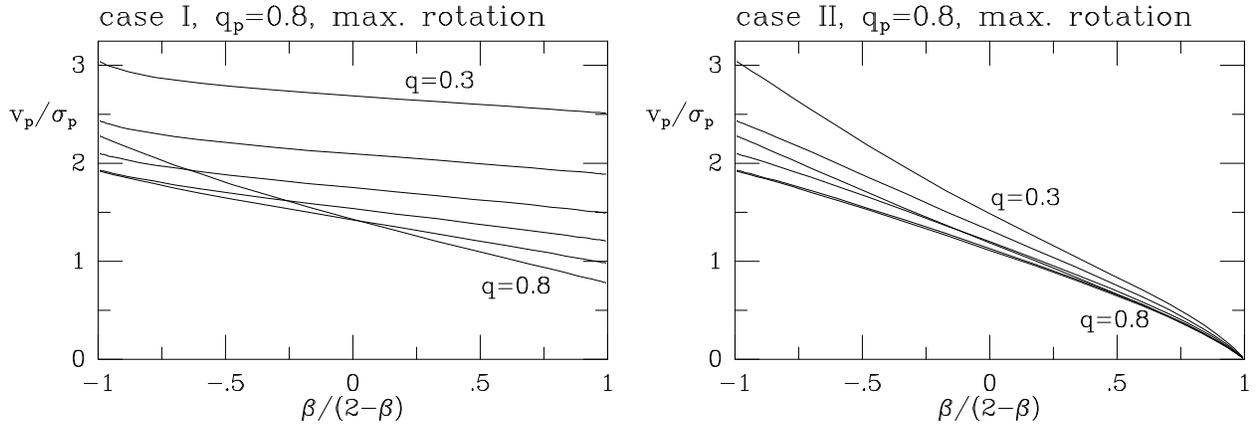}}
\caption{Ratio $v_{\rm p}/\sigma_{\rm p}$ of the mean
streaming and velocity dispersion on the projected major axis as
function of $\beta / (2-\beta)$, for the same set of models as in
Figure~4, using the maximally rotating odd part. The left edge of each
panel corresponds to $\beta \rightarrow -\infty$, the right edge to
$\beta \rightarrow 1$. Less streaming is possible in models with large
$\beta$, i.e., in models with more radial motion.}
\end{figure*}

\subsection{Velocity profiles}

The Gauss--Hermite coefficients that characterize the VP shapes of the
models can be calculated as described in Section~2.5 and Appendix~A.
As an example, consider the model with $\gamma=4$, $\delta=1$, axial
ratio $q=0.8$ and inclination $i=90^{\circ}$. Based on the results of
the previous section, the discussion is restricted to the case~II
DFs. The parameter $s$ of the odd part is varied from ${1\over2}$ to
1, while $t$ is set to zero (i.e., $f_{\rm o}$ is equal to $f_{\rm e}$
times a step function, cf.~eq.~[\ref{df_odd}]). This yields models
that range from non--rotating to maximally rotating. Figure~6 shows
the Gauss--Hermite coefficients $h_3$ and $h_4$ for different values
of $\beta$. The abscissa in the figure is the observationally
accessible quantity $v_{\rm p} / \sigma_{\rm p}$, which increases
monotonically with the model parameter $s$.  The predicted
Gauss--Hermite coefficients depend in a complicated way on the model
parameters $\gamma$, $\delta$, $q$ and $i$, but none the less, the
results in Figure~6 are generic for a wide variety of parameter
combinations.

As argued in Section~3.3, the models that best fit real galaxies are
probably those with $\beta \ga 0$ in which the rotation is
significantly less than the maximum possible. Figure~6 shows that
these models predict $-0.1 \la h_4 \la 0.1$. Opposite signs are
predicted for $h_3$ and $v_{\rm p} / \sigma_{\rm p}$, provided that
$\beta$ is not too close to unity. These predictions agree well with
the observations of nearly all galaxies for which VP information is
available (e.g., van der Marel \& Franx 1993; van der Marel \etal
1994a; Bender \etal 1994).

The even part of the VP is fully determined by the even part of the
DF, and hence is independent of either $s$, $t$ or $v_{\rm p} /
\sigma_{\rm p}$. The fourth-order Gauss--Hermite
moment of this even part is sometimes referred to as $z_4$ (e.g., van
der Marel \etal 1994b). In Figure~6 its value is read off as the value
of $h_4$ at $v_{\rm p} / \sigma_{\rm p} = 0$.

\begin{figure*}
\centerline{\epsfbox{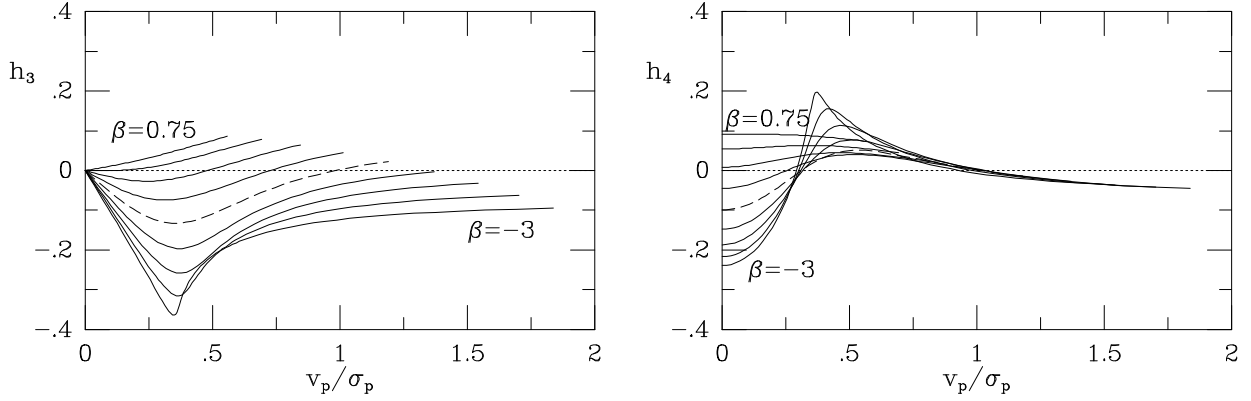}}
\caption{The Gauss--Hermite coefficients $h_3$ and $h_4$ on the projected 
major axis, as function of the ratio $v_{\rm p}/\sigma_{\rm p}$ of the
mean streaming and velocity dispersion. These coefficients measure
deviations of the VPs from a Gaussian. The model has a Kepler
potential ($\delta=1$), a mass density with power-law slope $\gamma=4$
and axial ratio $q=0.8$, and is viewed edge--on. The curves are for
models with a case~II DF, with $\beta=-3.00$, $-1.83$, $-1.00$,
$-0.41$, $0$ (dashed curves), $0.29$, $0.50$, $0.65$ and $0.75$,
respectively. The odd part of the DF has $t=0$, while the parameter $s$
is varied to produce models with different $v_{\rm p}/\sigma_{\rm p}$,
ranging from non--rotating to maximally rotating.}
\end{figure*}

\section{An application}

As an application of the models, consider the issue of the dynamical
detection of dark matter in elliptical galaxies. Both tangential
anisotropy and the presence of a dark halo can cause the observed
velocity dispersion ${\sigma}_{\rm p}$ to remain roughly constant as
function of galactocentric distance $R'$, out to well beyond the
effective radius $R'_{\rm eff}$. Proving the presence of a dark halo
therefore requires the construction of anisotropic axisymmetric
models, to rule out the possibility of strong tangential anisotropy.

Carollo \etal (1995) presented stellar kinematical data for the four
elliptical galaxies NGC 2434, 2663, 3706 and 5018, out to $\sim 2
R'_{\rm eff}$. Here the discussion is restricted to two of these
galaxies, NGC 2434 and NGC 3706. Carollo \etal interpreted their data
by constructing flattened models with $f=f(E,L_z)$. From combined
modelling of the major axis rms projected line-of-sight velocity
$\langle v_{z'}^2 \rangle_{\rm p} = \sigma_{\rm p}^2 + v_{\rm p}^2$,
and the Gauss--Hermite coefficient $z_4$ they concluded that the data
for neither galaxy can be fit by any DF without invoking the presence
of a massive dark halo. They also showed that the dark halos must be
flattened, if the observed VPs are to be fit with an $f=f(E,L_z)$ DF.
 
The families of DFs presented here can be used to further interpret
the observations for NGC 2434 and NGC 3706. The model parameter
$\gamma$ is chosen to fit the observed surface brightness slope at
large radii, and $q_{\rm p}$ is chosen equal to the average apparent
flattening of the isophotes outside half the effective radius. This
yields $\gamma = 2.94$ and $q_{\rm p}=0.92$ for NGC 2434, and $\gamma
= 3.36$ and $q_{\rm p}=0.65$ for NGC 3706. The potential is chosen to
be logarithmic ($\delta = 0$), since it has already been demonstrated
that both galaxies must be embedded in a dark halo. This yields a flat
${\sigma}_{\rm p}$ profile. Following Carollo \etal we study the
Gauss--Hermite coefficient $z_4$. This quantity is fully determined by
the even part of the DF, and the only free parameters of the models
are thus $\beta$ and the inclination angle $i$.

Figure~7 shows the model predictions as function of the assumed
inclination. The data points at large radii fall between the two
horizontal dotted lines (these represent the error bars at the
outermost measured points, and are a conservative estimate of the
observational uncertainty in the mean $z_4$ for $R' \geq R'_{\rm
eff}$; additional systematic errors in the spectral analysis due to
template mismatching or continuum subtraction are believed to be at
most $\vert \Delta z_4 \vert \la 0.03$). The dashed curve shows the
predictions for the $f(E,L_z)$ model, i.e., case~I and $\beta=0$. The
predicted $z_4$ values fall below the observations for both galaxies,
as shown already by Carollo et al. They demonstrated in addition that
$f(E,L_z)$ models with a flattened dark halo do predict the correct
$z_4$ for both galaxies.  The required flattening of the dark matter
density is $q \approx 0.7$ for NGC 2434 and, somewhat implausibly, $q
\la 0.3$ for NGC 3706. The results of Figure~7 show that the data can
also be fit with a spherical dark halo and a more general DF. The
solid curves show the predictions for the case~II DFs with different
values of $\beta$. For both galaxies there is a significant range of
inclinations and values of $\beta$ for which the observed $z_4$ values
are matched. Hence, the VP shapes can be fit with both flattened and
spherical dark halos.  More data, e.g., along the minor axis, are
required to distinguish between the various possible models.

\begin{figure*}
\centerline{\epsfbox{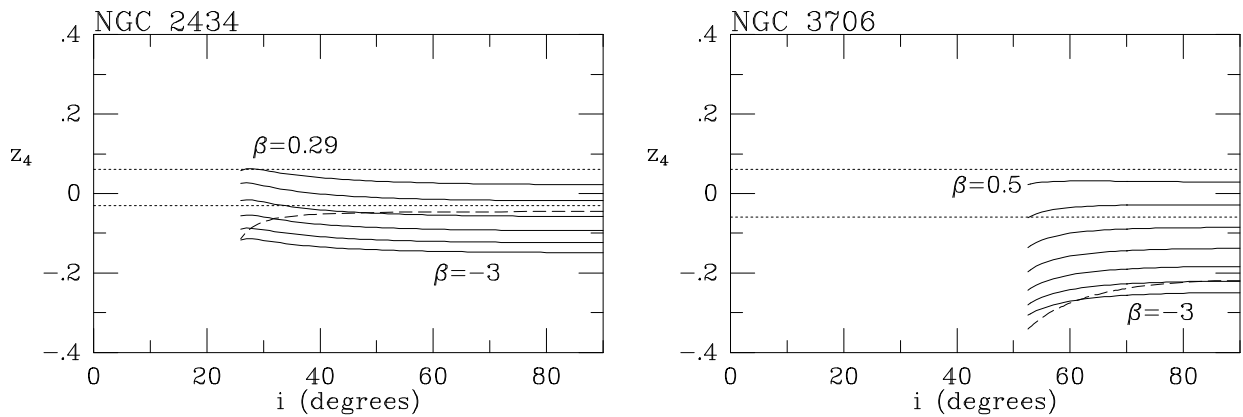}}
\caption {The Gauss--Hermite coefficient $z_4$ of the major axis VPs of
NGC 2434 and NGC 3706. This coefficient quantifies the lowest-order
deviations of the even part of the VP from a Gaussian. A value $z_4<0$
indicates that this even part is more flat--topped than a Gaussian,
and a value $z_4>0$ that it is more centrally peaked. The data points
outside the effective radius fall between the horizontal dotted
lines. The curves show the model predictions as function of the
assumed inclination. The lowest inclination plotted corresponds to an
intrinsic axial ratio $q=0.3$. Dashed curves are for $f(E,L_z)$
models, i.e., case~I and $\beta=0$. These models do not fit the
data. Solid curves are for case~II DFs and various values of $\beta$
($\beta=-3.00$, $-1.83$, $-1.00$, $-0.41$, $0$ and $0.29$, and in the
right panel also $\beta=0.5$). For these models a range of
inclinations and $\beta$ values provides an acceptable fit to the
data. Our algorithm for calculating $z_4$ (Appendix~A) does not work
well for very large values of $\beta$ in a logarithmic potential, but
calculations in other potentials indicate that $z_4$ keeps increasing
monotonically with $\beta$ for $\beta \rightarrow 1$.}
\end{figure*}

\section{Discussion and conclusions}

A study has been presented of stellar dynamical models for scale-free
flattened spheroidal mass densities in scale-free spherical
potentials. The mass density is characterized by its power-law slope
$\gamma$ and its flattening $q$, while the potential is characterized
by its power-law slope $\delta$. The general form of the DFs was
derived, and two particular families of DFs were studied in
detail. The DFs of these families are separable functions of the
integrals of motion, or combinations thereof. Both families have a
free parameter $-\infty < \beta < 1$ which regulates the velocity
dispersion anisotropy. In the spherical limit they reduce to
constant-anisotropy models of the type discussed by H\'enon
(1973). The DFs of the models can be expressed in terms of generalized
hypergeometric functions or power series with known coefficients,
which reduce to elementary functions in many cases of interest
(Tables~1 and 2). Because of their simple structure, it is relatively
straightforward to calculate the intrinsic and projected velocity
moments. The latter can be used to reconstruct the projected VP
shapes.

The `case~I' distribution functions are anisotropic generalizations of
the flattened $f(E,L_z)$ model, which they include as a special
case. The `case~II' distribution functions generate flattened
constant-anisotropy models. For both families, Binney's function
$\beta_{\rm B}$ on the minor axis is equal to the DF parameter
$\beta$. For the case~I DFs, the ratio $\langle v_{\rm t}^2 \rangle /
\langle v_r^2 \rangle$ increases monotonically with $\theta$. For the
case~II DFs it is independent of $\theta$, and the function
$\beta_{\rm B}$ is everywhere equal to the parameter $\beta$.  The
case~I and case~II DFs are identical in the limit $\beta \rightarrow
-\infty$, which is the model built exclusively with circular
orbits. Case~I DFs with $\beta=0$ correspond to the classical
$f(E,L_z)$ DF. Case~II DFs with $\beta \rightarrow 1$ correspond to
models with all stars on radial orbits. Calculations of the ratio of
the rms projected velocity on the projected major and minor axes show
that real elliptical galaxies are probably best described by the
case~II DFs with $\beta \ga 0$. Such models also predict VP shapes
consistent with observations.

Two important conclusions can be drawn: (i) flattened axisymmetric
stellar systems can have a large range of physical DFs and dynamical
structures; and (ii) only a small subset of the possible dynamical
structures is realized in nature. This agrees with the work of Dehnen
\& Gerhard (1993), who constructed self-consistent three-integral 
DFs for a flattened isochrone model. They restricted themselves to
`quasi-separable' functions of the integrals of motion, while the
present paper has been restricted to two special families of DFs. The
full range of possible DFs for flattened systems is therefore even
larger than that discussed in either paper.

As an application, the models are used to interpret the VP data
obtained recently by Carollo \etal (1995). They showed that the
galaxies NGC 2434 and NGC 3706 must have dark halos, and that the dark
halos must be flattened if the observed VPs are to be fit with an
$f=f(E,L_z)$ DF. Our models demonstrate that the data can be fit
equally well with a spherical dark halo, provided that the DF is more
general than $f=f(E,L_z)$. In particular, the case~II DFs with $\beta
\ga 0$ provide a good fit. Data along more position angles are
required to discriminate between the various possible models and dark
halo shapes.

A disadvantage of the models discussed here is that the potentials of
real galaxies are generally not spherical, especially not in the inner
regions, where the potential is dominated by the luminous matter. This
introduces a number of systematic differences with respect to the
predictions of self-consistent models. For example, the velocity
ellipsoids of the models always align with spherical coordinates,
whereas this need not be the case in self-consistent flattened models,
although it often is a good approximation (Dehnen \& Gerhard 1993; de
Zeeuw, Evans \& Schwarzschild 1995). Also, the tensor virial theorem
dictates that a flattened mass density in a flattened potential has
more rms motion parallel to the equatorial plane than the same mass
density in a spherical potential. On the other hand, the potentials
generated by flattened mass distributions are always more nearly
spherical than the density distribution itself, especially at large
radii where the monopole component of the potential dominates. Indeed,
the models illustrated in Figures~1 to~6 (Kepler potential, mass
density power-law slope $\gamma=4$ and flattening $q=0.8$) are the
asymptotic large-radii limit of the self-consistent models studied by
Dehnen \& Gerhard (1993). The potentials of real galaxies are probably
dominated by dark halos, at least in the outer parts, and these may
well be nearly spherical.

One case where the present models are certainly applicable is to the
central density cusp structure around a nuclear black hole, where the
potential is known to be spherical and Keplerian. From
equation~(\ref{DFpos}) with $\delta=1$ it follows that the physical
DFs must have $\beta < \gamma - {1\over2}$, for either the case~I or
case~II. So at least for the particular families of DFs studied here,
the presence of a shallow density cusp ($\gamma < {3\over2}$) around a
central black hole, precludes a large radial velocity dispersion
anisotropy. Conversely, the constraint on $\beta$ implies that oblate
$f(E,L_z)$ models (i.e., case~I, $\beta=0$) are physical only if
$\gamma >{1\over2}$, as shown previously by Qian \etal (1995).

A useful generalization of the present work would be to build triaxial
models in spherical potentials. This can be achieved by using DF
components that involve powers of $L_x^2$, $L_y^2$ and $L_z^2$, rather
than just $L^2$ and $L_z^2$ (Mathieu \etal 1995; Evans, private
communication).

\section*{Acknowledgments}

We thank Marcella Carollo for kindly providing the data on NGC 2434
and NGC 3706. RPvdM was supported by NASA through a Hubble Fellowship,
\#HF--1065.01--94A, awarded by the Space Telescope Science Institute
which is operated by AURA, Inc., for NASA under contract NAS5--26555.
PTdZ gratefully acknowledges the hospitality of the Institute for
Advanced Study in Princeton, and partial support from NSF grant PHY
92--45317.

\appendix

\section{Reconstruction of a velocity profile from its moments}

Various approaches exist for the recovery of a VP (\ref{VPdef}) from
its moments (\ref{losmom}).  One possibility is to reconstruct the VP
from its Gram--Charlier series expansion (e.g., van der Marel \& Franx
1993; Magorrian \& Binney 1994). This works well, but only for VPs
that are close to a Gaussian. In the present context it did not always
produce satisfactory results. An alternative approach was therefore
used, in which the VP is represented by its value on a discrete array
of $N$ velocities. The VP values are then estimated from the first $N$
velocity moments upon inversion of the so--called VanderMonde matrix
(e.g., Press \etal 1992, section~2.8). The observable quantities are
easily calculated from the discretized VP. This process was performed
iteratively, using higher and higher $N$, until convergence was
achieved in $(\gamma_{\rm G},V,\sigma,h_3,h_4)$. This generally works
well, despite the ill-conditioned nature of VanderMonde matrix
inversion. The reason for this is that the Gauss--Hermite series bears
a strong resemblance to a Fourier series.  Gauss--Hermite moments
$h_i$ of higher order measure power in the VP on higher frequencies
(Gerhard 1993). For high $N$, numerical errors cause the VanderMonde
solution to become oscillatory on the scale of the velocity array
spacing. This, however, does not influence the observables
$(\gamma_{\rm G},V,\sigma,h_3,h_4)$, which only measure power on
relatively low frequencies.

The `VanderMonde algorithm' was tested in various ways. A program was
written that calculates the VP for the spherical ($q=1$) case of the
models by direct evaluation of the three-dimensional integral in
equation~(\ref{VPdef}). The $(\gamma_{\rm G},V,\sigma,h_3,h_4)$
calculated with the VanderMonde algorithm were found to be accurate,
except for the case of a logarithmic potential and large anisotropy
($\beta \la -4.0$ or $\beta \ga 0.8$). This has two reasons: (i) the
logarithmic potential has no escape velocity so that it is more
difficult to represent the VP on a finite velocity array; and (ii) the
VP becomes discontinuous in the limit $\beta \rightarrow -\infty$
(only circular orbits), and singular in the limit $\beta=1$ (only
radial orbits) (e.g., van der Marel \& Franx 1993). A second test was
provided by the $f(E,L_z)$ case ($\beta=0$), for which the
three-dimensional VP integral~(\ref{VPdef}) could be calculated as in
Qian \etal (1995). Again, the $(\gamma_{\rm G},V,\sigma,h_3,h_4)$
calculated with the VanderMonde algorithm were generally found to be
accurate, with the exception of very flattened models $q \la 0.6$ in a
logarithmic potential. This is understood from the fact that the VPs
of these models become contrived and double--peaked for large
flattening (Dehnen \& Gerhard 1994). In both test cases inaccurate
results were accompanied by non-zero values of $h_1$ and $h_2$, which
should be zero by definition. For the general case we therefore took
the values of $h_1$ and $h_2$ as an indicator of the numerical
accuracy of the VanderMonde algorithm. From this it was found that the
algorithm fails only for the case of a logarithmic potential, large
flattening and large anisotropy. No attempt was made to develop an
algorithm to recover the VP in a more sophisticated way. This probably
requires some form of regularization of the problem (see, e.g.,
Merritt \& Tremblay 1994), which is outside the scope of this paper.

\bsp

\end{document}